\documentclass[letterpaper,pra,showpacs,showkeys,10pt,twoside,byrevtex,nofootinbib, superscriptaddress, amsfonts,amsmath]{revtex4-1}
 \usepackage{graphicx}
  \usepackage{amssymb}
  \usepackage{color}
  \begin{document}
\title{Quantum Mechanical Inclusion of the Source in the Aharonov-Bohm Effects}
\author{Philip Pearle}
\email{ppearle@hamilton.edu}
\affiliation{Emeritus, Department of Physics, Hamilton College, Clinton, NY  13323}
\author{Anthony Rizzi}
\email{arizzi@iapweb.org}
\affiliation{Institute for Advanced Physics, PO Box 15030, Baton Rouge, Louisiana, 70895}
\date{\today}
\begin{abstract}   

Following  semiclassical arguments by Vaidman we show, for the first time in a fully quantum mechanical way, that the phase shifts arising in the Aharonov Bohm (A-B) magnetic or electric effects can be treated  as due to the electric force of a classical electron,  respectively acting on quantized solenoid particles  or quantized capacitor plates.  This is in contrast to the usual approach which treats both effects as arising from non-field producing potentials acting on the quantized electron. Moreover, we consider the problems of interacting quantized electron and quantized solenoid or quantized capacitor  to see what phase shift their joint wave function acquires. We show,  in both cases, that the net phase shift is indeed the A-B shift (for, one might have expected twice the A-B shift, given the above two mechanisms for each effect.) The solution to the exact Schrodinger equation may be treated (approximately for the magnetic A-B effect, which we show using a variational approach,  exactly for the electric A-B effect) as the product of two solutions of separate Schrodinger equations for each of the two quantized entities, but with an extra phase. The extra phase provides the negative of the A-B shift, while the two separate Schrodinger equations without the extra phase each provide the A-B phase shift, so that the product wave function produces the net A-B phase shift.  \end{abstract}
\maketitle

\section{Introduction}

The Aharonov-Bohm (A-B) magnetic effect \cite{AB,FR, AR} predicts relative phase shifts of two electron wave packets moving in alternative paths around the outside of a long cylindrical solenoid.  Classically, it appears impossible that there should be a measurable effect in such circumstances, where the electron moves in a region of non-field producing potentials, i.e., in a region where there is a vector and/or scalar potential, but no electric or magnetic fields.  That such potentials can produce physical effects is considered by many to be a prime example of the marvelous novelty quantum theory reveals vis-a-vis classical theory. 

Recently, Vaidman \cite{Vaidman}  argued, using a semi-classical analysis of a model of the solenoid that, under the influence of the electric-field-producing vector potential of the electron treated classically, the solenoid  produces a relative phase shift (depending upon the direction of traverse of the electron wave packet)  exactly equal to the standard magnetic A-B phase shift.  
 
The importance of the A-B effect suggests that Vaidman's result should be carefully considered in the context of a fully quantum mechanical treatment of the solenoid particles.  We do this in this paper and  obtain Vaidman's result. 
 
However, given the radical interconnectedness implied by quantum mechanics, it is natural to consider the problem where \textit{both} the electron and solenoid 
particles are quantized\cite{PTT}.  Does one then get twice the A-B phase shift, the sum of the usual and Vaidman results? The answer is no: we consider that problem here, and find 
that, indeed, one gets the usual A-B phase shift. This occurs through a consistent treatment of the interaction between the particles of the solenoid and the electron. 

We model the solenoid as consisting of $N$ particles.  They, and the electron are considered to be described by well-defined wave packets, and the joint wave function is approximated as the product of such wave functions. Putting this approximate wave function into the variational principle for the Schr\"odinger equation results in $N+1$ Schr\"odinger equations, where 
each particle evolves under the vector potential of the other particle's mean positions and momenta, which are the positions and momenta  of the same classical problem.

However, there is an extra phase term in each Schr\"odinger equation. By a phase transformation which does not change the overall phase of the wave function, 
the extra phase term may be removed from all Schr\"odinger equations but one.  We choose to put that extra phase term into the electron's Schr\"odinger equation. That extra phase turns out to be 
the time integral of the interaction term in the Hamiltonian, when operators are replaced by their classical counterparts. We then suppose 
the $N+1$ particles are involved in an interference experiment, and show that the extra phase 
is the negative of the A-B phase shift.  The Schr\"odinger equation for the electron (evolving in the classical field of the solenoid), without the extra phase term, 
gives the usual A-B phase shift.  Therefore, the Schr\"odinger equation for the electron with the extra phase term gives 0 phase shift.  The Schr\"odinger equation for the $N$ 
solenoid particles gives the A-B phase shift. Thus, the net result for the product wave function is the A-B phase shift.

Although we do not bother to consider it here, the extra phase might just as well have been eliminated from the electron's Schr\"odinger equation 
and distributed in any way among the $N$ solenoid particles.  The result then is that the joint wave function of the solenoid particles, including the extra phases, produces zero 
phase shift in the interference experiment.  Then, the A-B phase shift is attributable solely to the electron's wave function, as in the usual description.
 
In a companion paper\cite{PR2}, we point out that consideration of \textit{all} relevant quantum objects in this problem requires quantizing the vector potential as well. 
A similar analysis of the joint wave function of electron, solenoid and vector potential, approximated as a product wave function and 
governed by the separate Schr\"odinger equations which arise from the variational principle, gives the usual A-B phase shift. As in this paper, depending upon which 
Schr\"odinger equation(s) get the extra phase, the A-B shift  may be attributed solely to the wave functon of the electron, or solely to the solenoid, or additionally solely to the vector potential, or 
to a combination thereof.

Our analysis also shows why the usual phase shift due to the electron, and the phase shift due to the solenoid have to be the same.  To complete 
the story we spell out the details of the model of the solenoid, show that the amplitude of the interference term is essentially 1 for reasonable values of the model parameters, and do the direct calculation of the A-B phase shift arising from the solenoid.
 
Next, we turn to show that similar considerations apply in the case of the electric A-B effect.
Vaidman  proposed a simplified physical setup wherein the electric potential of two classical charges produces a zero-field environment for the electron and, thereby, an electric A-B phase shift.  
Then, he argued for an alternative treatment, where the electron is treated classically and the charges are quantized, and showed that the electric 
field of the electron produces the same electric A-B phase shift for the two charges. 

We shall consider here, not Vaidman's simplified model, but  the standard A-B electric phase shift situation, of an electron passing over or under a charged capacitor.  The usual calculation attributes the phase shift to the electron packets moving in 
regions of different constant scalar potential created by a classical capacitor.  We shall show,  as Vaidman did in his quasi-classical analysis for his model, that this same A-B phase shift 
is obtained when the electron is treated classically 
and the capacitor plates (more precisely, the plate's center of mass coordinates) are treated quantum mechanically, thereby acquiring the A-B phase shift in the electric field of the electron.  
Again, these are two alternative,  mathematically but not conceptually equivalent, ways to calculate the same thing.   So, we consider the situation where \textit{both} the electron and the capacitor plates have quantum dynamics. Similarly to the magnetic A-B effect, we show how the joint phase shift may be attributed solely to the electron's motion, or may be attributed solely to the capacitor plate motion, simply by shifting a term in the Hamiltonian to either the electron part of the Hamiltonian or to the capacitor plate part of the Hamiltonian, 
or that other splits may be made. 

Thus, for both these classic examples of the A-B effects, there is support for an alternative (espoused by Vaidman) to the usual views of 
 these effects as due to the electron moving in a non-field producing potential. It is that the effects may, with equal justification  be viewed 
 as due to motion of charged objects due to forces exerted by the electron.  However, other considerations than ours may lead one to prefer one point of view over the other.

The plan of this paper is as follows. 
   
 In Section II, we calculate the amplitude and phase of a wave packet describing a charged particle with well-localized position and momentum  on a one-dimensional pre-determined path 
(e.g., moving in a  tube with high potential walls),  under arbitrary time-varying (but, not space-varying) externally applied electric field-producing vector and scalar potentials, for a time interval $T$. We show that the phase can be completely expressed in terms of the classical motion of the particle (which is the mean motion of the wave packet).   
  
 In Section III,  we generalize the result of the previous section, to calculate the amplitude and phase associated to a wave packet in the more general case where the 
 external electromagnetic  forces experienced by the particle depend not only upon arbitrarily upon time, but also arbitrarily upon the location of the particle as well.  
 The approximate solution we develop is for the case in which the particle's mean position follows the classical path. 
 
 In Section IV,  we first consider two quantized particles interacting via their mutual vector and scalar potentials. As described above, we approximate 
 the wave function as the product of localized wave packets for each particle.  Using the variational principle 
 for the exact Schr\"odinger equation,  we derive Schr\"odinger equations for each particle. 
 We show the Schr\"odinger equations describe each particle as moving under the potentials of the 
 other particle treated classically.  This allows us to use the results of Section III to calculate the amplitude and phase of each particle.  But, in addition,
 each Schr\"odinger equation contains an extra phase term.  By adding a phase to one particle and subtracting it from the other, so the overall 
 phase of the wave function is not affected, the extra phase is removed from one particle and belongs solely the other. 
 
 It is shown that the extra phase is the time integral of the interaction term in the Hamiltonian, 
  where all operators have been replaced by the corresponding classical variables: call this $-\Phi$. It is shown that the 
  Schr\"odinger equation for the particle without the extra phase, and the other particle's Schr\"odinger equation 
  with the extra phase removed, each produce the same phase, $\Phi$. So, the total phase is $\Phi$.  And, in an interference experiment involving the two particles 
  moving on two different paths, it is the value of $\Phi$ on one path minus the value of $\Phi$ on the other path that provides the total phase shift.  
  
We then extend these results to the case of $N$ particles, each interacting with one ``special" particle we call the ``electron." 
    Again, the exact wave function is approximated as a product of all $N+1$ particle wave functions. 
    Schr\"odinger equations are obtained for each particle and each equation has an extra phase term which may be eliminated from the $N$ particles, 
    leaving the electron wave function with all the extra phase. 
    
The resulting phase situation is as follows.  The electron's extra phase term is $-\Phi$, so the electron  Schr\"odinger equation 
produces 0 phase. Then, the net phase is that of the $N$ particles, which is $\Phi$. 
 Again, in a considered interference experiment where there are two different paths for all particles, the total net phase shift is 
 the difference of this phase $\Phi$ for the two paths.
    
 Additionally, for this interference experiment, we obtain an expression for the amplitude of the 
  interference term, expressed as a function of the $N$ particle classical positions and momenta.
  
In Section V  we present a fully quantum mechanical approach to Vaidman's 
  semi-classical considerations.  In particular, we apply the argument given above to a detailed model 
of the solenoid as a collection of $N$ well-localized circulating quantized particles, interacting with the electron.  The results of the preceding section immediately apply. The total net phase shift, the difference of the net phase of the joint electron-solenoid wave function for two different paths taken by the electron, is the A-B phase shift. 

 Also, it is explained why the Vaidman scenario calculation \textit{must} produce
the same phase shift as the usual calculation.  For further insight, we give an intuitive time-averaged derivation of our result. 

Furthermore, we consider the magnitude of the interference term in this A-B experiment, which experimentally\cite{Ton} is 1. In Vaidman's semi-classical calculation of the magnetic A-B effect, it is necessary for the electron at the end of its traverse to 
 decelerate back to zero speed and thus cancel the initial impulse that started the solenoid cylinders moving.  This is in order that each cylinder's final motion be unchanged from its initial (pre-acceleration) motion and its entanglement with the electron 
packets be removed, else there would be no interference\cite{Vaidman}.  In our case, there is no need for such a caveat.  We show in our model, despite the two different displacements of each solenoid particle associated with the two different electron paths, that the magnitude of the interference term is close to 1 because the displacements are very small 
(since  each contributes only a very tiny amount to the phase shift.)

This completes our presentation of the A-B magnetic effect, showing how the A-B phase shift with interference amplitude 1 is obtained when considering both the quantized electron and quantized solenoid. 

 Section VI applies the result of Section III to the electric A-B effect.  

We show that the electric case is exactly parallel to the magnetic A-B case.  We 
consider the exact problem wherein we have a quantized electron and quantized capacitor plates (modeled as sheets of glued-together charges, so their centers of mass 
are the plate dynamical variables).   
We  explain how the phase shift expression can either be written as if it were due to the electron moving in the non-field producing potential external to the capacitor or, alternatively, as if it were due to the capacitor plates moving in the electric field of the electron, or in some other equivalent way.   

Section VII contains some concluding, summary remarks. 

 We set $\hbar=1$.  In order to avoid superfluous $4\pi$ factors, we employ cgs units in all but Section VI and Appendix D, where we employ rationalized MKS units.    

\section{Particle Moving Under Time-Dependent Non-Spatially Varying Forces.}

In this and the next section, we consider the general problem of a particle well-localized in position and momentum and moving in one dimension under scalar and vector potentials,  and calculate the magnitude and phase of its wave function. 
 
The problem of a particle moving in one dimension is to be considered as a limiting case of the three dimensional problem of a particle moving in 
 a three dimensional ``pipe." This is relevant here, because we shall model the solenoid as a collection of particles moving in stacked tubes, i.e., the ``pipe"  is a tube of toroidal shape. A bonus is that we can treat the orbiting ``electron"  the same way, taking its motion to be  in a circular tube
 of its own, concentric with and outside the solenoid.
 
 In cylindrical coordinates, the kinetic energy in the Hamiltonian is separable, and so a tube may be modeled as providing, say, a square well potential in the radial and $z$-directions, and supposing that the particle  under consideration is 
 in the ground state in these directions.  Thus,  it has free motion in the azimuthal direction.   If the sides of the tube are allowed to be as small as desired, the separation in energy between the ground and lowest excited states gets as large as desired, so that any external field has negligible probability of exciting the radial or $z$-motion, and just consists in a distortion of the ground state.  Thus, the only interesting motion is the free, one-dimensional particle motion. 
 
Of course, this one dimensional motion approximation to describing a linearly constrained three dimensional motion has greater applicability than to 
 just circular motion in cylindrical coordinates.  Immediately, one notes that this argument is just as readily applied to motion along a constant parameter curve in any of the 13 
 coordinate systems in which the Laplacian is separable.  It should also be applicable to motion along a larger class of curves, along which one can choose 
 local coordinates so that the Laplacian is effectively separable, at least into  the one-dimensional linear motion and two-dimensional orthogonal motion: we shall not pursue here an analysis of the  class of curves for which 
 the one-dimensional motion approximation to constrained three dimensional motion has validity.
 
Obviously, this is just one possible configuration for the magnetic A-B effect.  In the only actually rigorously performed experiment\cite{Ton}, the solenoid was replaced by a small magnetized torus and, after  one packet goes through the torus hole and the other passes outside, the electron wave functions were allowed to spread and overlap. 
While the results obtained here really only apply to the configuration of our model, they suggest that there \textcolor{red}{should} be counterparts when other configurations are considered.

In this section, we consider that the motion of the particle in one dimension is under scalar and vector potentials associated to forces which are solely time-dependent (i.e., not space-independent).  This  is an exactly solvable problem, and we find the expression for the amplitude and phase acquired by the particle's wave function in this case.  We do this before tackling  in the next section the more general, but not exactly solvable problem where the forces depends upon the particle's position as well as upon time.  

The result from this section, limited to solely time-dependent (non-spatially varying) forces,  is however applicable in Section VI, to the case of the quantized capacitor plates moving under the electron's 
classical Coulomb field.  The plates are assumed to be so massive that they do not 
displace significantly during the electron's traverse, which simplifies the calculation.

\subsection {Classical motion}

First, consider  the \textit{classical} motion of a particle, of charge $q$ and mass $m$, under solely time-dependent electric forces due to a vector and scalar potential.  

If the vector potential is  $A(t)$, then
its associated electric force $-q\frac{d}{cdt}A(t)$ has just time dependence.  

Expand the potential energy $qV(x,t)$ for a particle which moves on the classical trajectory $x=x_{cl}(t)$ 
about the particle's position:  $qV(x,t)=qV(x_{cl}(t),t)+(x-x_{cl}(t))qV'(x_{cl}(t),t)+...$, where $x$ is the distance along the path.  Since  the force felt by the particle is 
to have no spatial dependence, we assume no higher power of $(x-x_{cl}(t))$ than the first.  Moreover, since all the problems of concern to us are of first order in $q$, and since 
$x_{cl}(t)=x_{0}+v_{0}t+o(q)$ ($x_{0}$ is the particle's initial position, $v_{0}$ its initial velocity), we replace $x_{cl}(t)$ by $x_{0}+v_{0}t$.  Thus, we write the scalar potential energy 
 we shall consider as $qV(t)+(x-x_{0}-v_{0}t)qV'(t)\equiv qg(t)+qxV'(t)$ (where $g(t)\equiv V(t)-(x_{0}+v_{0}t)V'(t)$), i.e., we define $V(t)\equiv V(x_{cl}(t),t)$ and $V'(t)\equiv V'(x_{cl}(t),t)$.  

The classical Hamiltonian we consider is therefore 
\begin{equation}\label{1}
H=\frac{1}{2m}[p-\frac{q}{c}A(t)]^{2}+qxV'(t) +qg(t), 
\end{equation}
\noindent where $x$, $p$ are the canonical coordinates. 

The equations of motion follow from Hamilton's Poisson bracket equations:
\begin{eqnarray}\label{2}
\frac{d}{dt}x&=&\frac{1}{m}[p-\frac{q}{c}A(t)]\nonumber\\
\frac{d}{dt}[p-\frac{q}{c}A(t)]&=&-qV'(t)-\frac{q}{c}\frac{d}{dt}A(t) 
\end{eqnarray}
\noindent and solution 
\begin{subequations}\label{3}
\begin{eqnarray}
v_{cl}(t)&=&v_{0}-\frac{q}{mc}[A(t)-A(0)]-\frac{q}{m}U(t),\label{3a}\\
x_{cl}(t)&=&x_{0}+v_{0}t-\frac{q}{mc}[{\cal A}-A(0)t]-\frac{q}{m}[tU(t)-W(t)],   \\                       
p_{cl}(t)&=&mv_{cl}(t)+\frac{q}{c}A(t) =p_{0}-qU(t)  \\            
\hbox{where }{\cal A}&\equiv&\int_{0}^{t}dt'A(t')\\
\int_{0}^{t}dt'\int_{0}^{t'}dt''V'(t'')&=&\int_{0}^{t}dt'[t-t']V'(t')=tU(t)-W(t),\\
 p_{0}&\equiv&mv_{0}+\frac{q}{c}{A}(0),\quad U(t)\equiv \int_{0}^{t}dt'V'(t'), \quad W(t)\equiv \int_{0}^{t}dt't'V'(t').
\end{eqnarray}\label{3}
\end{subequations}
\noindent (The subscript $cl$ is to distinguish these classical variables from the quantum variables to appear later.)
 
Note, the electric force only appears in these equations:  a magnetic force does not appear for motion constrained to one dimension since that force is in the direction perpendicular to the velocity. 

\subsection{Wave function}

The wave function in the momentum representation satisfies
\begin{equation}\label{4}
i\frac{\partial}{\partial t}\psi(p,t)=\bigg[\frac{1}{2m}[p-\frac{q}{c}A(t)]^{2}+qg(t)+qV'(t)i\frac{\partial}{\partial p}\bigg]\psi(p,t), 
\end{equation}
(we set $\hbar=1$.)  It is initially taken to describe an object of momentum $p_{0}$ located at time 0 at $x_{0}$:
\begin{equation}\label{5}
\psi(p,0)=N e^{-(p-p_{0})^{2}\sigma^{2}}e^{-ipx_{0}}.
\end{equation}

To solve Eq.(\ref{4}) with the initial condition (\ref{5}), we make the ansatz
\begin{equation}\label{6}
\psi(p,t)=Ne^{-(\sigma^{2}+it/2m) p^{2}+\beta(t) p+i\gamma(t)-\beta_{R}^{2}(t)/4\sigma^{2}}, 
\end{equation}
\noindent where  $\beta=\beta_{R}+i\beta_{I}$, and $\gamma$ is real.  The last factor in the exponent of (\ref{6}) ensures $\int dp|\psi(p,t)|^{2}=N^{2}$.  From (\ref{5},\ref{6}), the initial conditions are $\gamma(0)=0$ and $\beta(0)=2p_{0}\sigma^{2}-ix_{0}.$

In Appendix \ref{A} we put  (\ref{6}) into (\ref{4}) and equate coefficients of $p$ and 1, with the results 
\begin{eqnarray}\label{7}
\beta_{R}&=&2\sigma^{2}p_{cl}(t), \quad \beta_{I}=-x_{cl}(t)+\frac{p_{cl}(t)t}{m},\quad \gamma=-q\int_{0}^{t}dt'V(t')+v_{0}W(t).
\nonumber\\
\end{eqnarray}

We then go to the position representation by taking the Fourier transform of (\ref{6}).  The result, (\ref{A12}), is 
\begin{equation}\label{8}
\psi(x,t)=Ne^{-\frac{(x-x_{cl}(t))^{2}}{4\sigma^{2}}}e^{ip_{cl}(t)(x-x_{cl}(t))}e^{ip_{cl}^{2}(t)t/2m}e^{-iq\int_{0}^{t}dt'V(t')+iv_{0}W(t)}.
\end{equation}

\noindent under the assumptions that the mass of the object is large enough so 
that there is negligible spreading of the wave packet over time $t$ and that the packet width $\sigma$ is much larger than the wavelength associated to the momentum, 
$\sigma>>h/mv$, so the momentum is well-defined.

Thus we see that it is a consequence of quantum theory that the mean particle motion follows the classical trajectory, and that 
the phase shift is expressed in terms of the classical variables as well.  

\section{Particle Moving Under General  Forces}

We now consider the motion of a charged particle in one dimension under externally applied potentials with arbitrary space and time dependence.  
An exact solution of Schr\"odinger's equation cannot be obtained for this more general case.  Instead, we resort 
to an approximate method (which, as we shall see, gives the same result for the phase as above when the force is restricted to be solely time-dependent).  
We set the stage for the approximation in subsections A and B. The approximation, consisting  of expanding the exact Hamiltonian 
about the classical trajectory of the particle, is presented in subsection C.  For a wave packet whose mean follows the classical trajectory, this is an exactly soluble problem, and the phase of the wave function is obtained in subsection D.

\subsection{Classical Motion}

	The motion  is governed by the classical Hamiltonian
\begin{equation}\label{9}
H=\frac{1}{2m}[p_{cl}-\frac{q}{c}A(x_{cl},t)]^{2}+qV(x_{cl},t)  
\end{equation}
\noindent where $x_{cl}$, $p_{cl}$ are the canonical coordinates.   

The equations of motion follow from Hamilton's Poisson bracket equations:
\begin{eqnarray}\label{10}
\frac{d}{dt}x_{cl}(t)&=&\frac{1}{m}[p_{cl}(t)-\frac{q}{c}A(x_{cl}(t),t)]\nonumber\\
\frac{d}{dt}[p_{cl}(t)-\frac{q}{c}A(x_{cl}(t),t)]&=&-qV'(x_{cl}(t),t)-\frac{q}{c}\dot A(x_{cl}(t),t)
\end{eqnarray}
\noindent where the dot means $\partial_{t}$ and the prime means $\partial_{x}$, and we have used $dF(x_{cl}, p_{cl}, t)/dt=[F,H]_{PB}+\partial F(x_{cl}, p_{cl}, t)/\partial t$..

Denoting ${\cal E}(x_{cl}(t),t)\equiv -V'(x_{cl}(t),t)-\frac{1}{c}\dot A(x_{cl}(t),t)$, the solution may formally be written 
\begin{eqnarray}\label{11}
v_{cl}(t)&\equiv&\frac{1}{m}[p_{cl}(t)-\frac{q}{c}A(x_{cl}(t),t)]=v_{0}+\frac{q}{m}\int_{0}^{t}dt_{1}{\cal E}(x_{cl}(t_{1}),t_{1}),\nonumber\\
x_{cl}(t)&=&x_{0}+v_{0}t+\frac{q}{m}\int_{0}^{t}dt_{1}\int_{0}^{t_{1}}dt_{2}{\cal E}(x_{cl}(t_{2}),t_{2}).
\end{eqnarray}

\subsection{Quantum Phase}

For the comparable quantum problem, the Hamiltonian is 
\begin{equation}\label{12}
H=\frac{1}{2m}[P-\frac{q}{c}A(X,t)]^{2}+qV(X,t) . 
\end{equation}
\noindent where $X,P$ are the conjugate operators. 
In this case we shall \textit{assume} the wave packet follows the classical trajectory, supposing that it  
spreads negligibly  over the time interval of interest  (i.e., $t/2m<<\sigma^{2}$):
\begin{equation}\label{13}
\psi(x,t)=e^{-[x-x_{cl}(t)]^{2}/4\sigma^{2}}e^{i\theta(x,t)},
\end{equation}
\noindent and see whether the phase $\theta(x,t)$ of the wave function can be obtained from Schr\"odinger's equation in some reasonable approximation.

As is well known from deBroglie-Bohm theory, putting (\ref{13}) into Schr\"odinger's equation results in two equations, the imaginary part yielding conservation 
of probability, in the form $\frac{\partial}{\partial t}\rho(x,t)+ \frac{\partial}{\partial x}[\rho(x,t)v(x,t)]$, the real part yielding the Hamilton Jacobi equation, modified by a``quantum potential:"
\begin{eqnarray}\label{14}
0&=&\frac{\partial}{\partial t}e^{-[x-x_{cl}(t)]^{2}/2\sigma^{2}}+\frac{\partial}{\partial x}\Big[e^{-[x-x_{cl}(t)]^{2}/2\sigma^{2}}\frac{1}{m}[\theta'(x,t)-\frac{q}{c}A(x,t)]\Big],\nonumber\\
-\dot\theta(x,t)&=&\frac{1}{2m}[\theta'(x,t)-\frac{q}{c}A(x,t)]^{2}+qV(x,t)-\frac{1}{2m}e^{[x-x_{cl}(t)]^{2}/4\sigma^{2}}\frac{\partial^{2}}{\partial x^{2}}e^{-[x-x_{cl}(t)]^{2}/4\sigma^{2}}.
\end{eqnarray}

The conservation equation, the first of (\ref{14}), becomes
\begin{equation}\label{15}
0=-\frac{1}{\sigma^{2}}[x-x_{cl}(t)]\Big[-v_{cl}(t)+\frac{1}{m}[\theta'(x,t)-\frac{q}{c}A(x,t)]\Big]+\frac{\partial}{\partial x}\frac{1}{m}[\theta'(x,t)-\frac{q}{c}A(x,t)],
\end{equation}
\noindent with solution
\begin{equation}\label{16}
v_{cl}(t)-\frac{1}{m}[\theta'(x,t)-\frac{q}{c}A(x,t)]=Ce^{[x-x_{cl}(t)]^{2}/2\sigma^{2}}.
\end{equation}
\noindent The expectation value of the momentum, $\int dx |\psi(x,t)|^{2}\theta'(x,t)$, will diverge unless $C=0$ (since the integrand goes $\sim C $ for large $x$), so the conservation 
of probability provides the condition 
\begin{equation}\label{17} 
\theta'(x,t)=mv_{cl}(t)+\frac{q}{c}A(x,t).
\end{equation}
The ``quantum potential" part of the second equation of (\ref{14}) is
\begin{equation}\label{18}
-\frac{1}{2m}e^{[x-x_{cl}(t)]^{2}/4\sigma^{2}}\frac{\partial^{2}}{\partial x^{2}}e^{-[x-x_{cl}(t)]^{2}/4\sigma^{2}}=-\frac{1}{8m\sigma^{4}}\Big[[x-x_{cl}(t)]^{2}-2\sigma^{2}\Big].
\end{equation}
\noindent Because the magnitude of the wave function keeps $[x-x_{cl}(t)]^{2}$ of the order of $\sigma^{2}$, (\ref{18}) is of magnitude $1/m\sigma^{2}$.  This may be neglected compared 
to the kinetic energy term in $(\ref{14})$, $mv_{cl}^{2}/2$, since we assume that there are many wavelengths within $\sigma$, so that the momentum of the packet is well-defined.  

Thus, we have the classical Hamilton-Jacobi equation, 
\begin{equation}\label{19}
-\dot\theta(x,t)=\frac{1}{2m}[\theta'(x,t)-\frac{q}{c}A(x,t)]^{2}+qV(x,t).
\end{equation}
In order that these two equations, (\ref{17}), (\ref{19}), be integrable, it must be that $\partial_{t}\partial_{x}\theta(x,t)=\partial_{x}\partial_{t}\theta(x,t)$.  From 
(\ref{17}) we obtain:
\begin{eqnarray}\label{20}
\partial_{t}\partial_{x}\theta(x,t)&=&\partial_{t}[mv_{cl}(t)+\frac{q}{c}A(x,t)]=\frac{d}{dt}mv_{cl}(t)+\frac{q}{c}\dot A(x,t)\nonumber\\
&=&-qV'(x_{cl}(t),t)-\frac{q}{c}[\dot A(x_{cl}(t),t)-\dot A(x,t)].
\end{eqnarray}\noindent where, in going from the first line to the second line in (\ref{20}), we have used the second of Eqs.(\ref{10}).
From (\ref{19}), with use of (\ref{17}), we obtain:
\begin{equation}\label{21}
\partial_{x}\partial_{t}\theta(x,t)=-\partial_{x}\frac{m}{2}v_{cl}^{2}(t)-qV'(x,t)=-qV'(x,t).
\end{equation}

We see that the right hand sides of (\ref{20}), (\ref{21})  are not, in general, identical, so these equations are not, in general, integrable.  

The reason is that we have assumed the wave packet to have a very specific trajectory, its center at $x=x_{cl}(t)$,  with specific initial conditions $x_{0}$ and $v_{0}$, 
whereas the Hamiltonian dynamics is general, not specific to this trajectory.   

While not integrable for this general case of the electric force depending upon position and time, 
they \textit{are} integrable in the special case earlier dealt with, where the electric field only depends upon time.  That is, if
$V'(x,t)=V'(t)$, $A(x,t)=A(t)$, the right hand sides of (\ref{20}), (\ref{21})  are identical.

\subsection{Classical Motion: Approximate Hamiltonian}

Our procedure shall be to construct an \textit{approximate} Hamiltonian  for which the problem is \textit{exactly} soluble (i,e, for which   the phase angle \textit{is} integrable). 

The Hamiltonian shall be designed so that the particular classical trajectory $x_{cl}(t)$, with specified initial values of $x_{0}, v_{0}$, which satisfies the exact Hamiltonian's equation of motion, now also satisfies the approximate Hamiltonian's   
equation of motion,  and that neighboring classical trajectories 
stay ``close" (to be made precise at the end of this section) to $x_{cl}(t)$.  
The Hamiltonian we propose is
\begin{equation}\label{22}
H'=\frac{1}{2m}[P_{cl}-\frac{q}{c}A(x_{cl}(t),t)]^{2}+qV(x_{cl}(t),t)+ (X_{cl}-x_{cl}(t))qV'(x_{cl}(t),t)-\frac{q}{c}v_{cl}(t)(X_{cl}-x_{cl}(t))A'(x_{cl}(t),t)
\end{equation}
 \noindent where now $X_{cl}, P_{cl}$ are the canonical coordinates. 
 
 Thus, one particular $x_{cl}(t)$ plays three roles: it is a solution of the exact Hamiltonian dynamics,  it is a crucial element in the approximate Hamiltonian (\ref{22}) and, as we shall see, 
 it is a solution of the approximate Hamiltonian dynamics.  
 
 The Poisson bracket equations of motion corresponding to the Hamiltonian (\ref{22}) are then
\begin{eqnarray}\label{23}
\frac{d}{dt}X_{cl}(t)&=&\frac{1}{m}[P_{cl}(t)-\frac{q}{c}A(x_{cl}(t),t)]\nonumber\\
\frac{d}{dt}[P_{cl}(t)-\frac{q}{c}A(x_{cl}(t),t)]&=&-qV'(x_{cl}(t),t)+\frac{q}{c}v_{cl}(t)A'(x_{cl}(t),t)-\frac{d}{dt}\frac{q}{c}A(x_{cl}(t),t) 
\end{eqnarray}
\noindent Upon combining these two equations, we obtain 
\begin{eqnarray}\label{24}
\frac{d^{2}}{dt^{2}}X_{cl}(t)&=&-qV'(x_{cl}(t),t)-\frac{q}{c}\dot A(x_{cl}(t),t).
\end{eqnarray}
 \noindent $X_{cl}(t)$ is the general solution with arbitrary initial conditions $X_{cl}(0),  V_{cl}(0)$.  We see from (\ref{10}) that  $X_{cl}(t)=x_{cl}(t)$ is a solution of (\ref{24}) (the solution with initial conditions 
 $X_{cl}(0)=x_{cl}(0), V_{cl}(0)=v_{cl}(0)$).  Moreover, $d^{2}[X_{cl}(t)-x_{cl}(t)]/dt^{2}=0$ so, for an arbitrary solution $X_{cl}(t)$, 
 \[
 X_{cl}(t)-x_{cl}(t)=[X_{cl}(0)+V_{cl}(0)t]-[x_{cl}(0)+v_{cl}(0)t]+ot^{3}.
\] 
Thus, two trajectories with the same initial speed and different initial positions only diverge to order $t^{3}$ under this Hamiltonian.    This property, of neighboring classical trajectories staying close, might be expected of a soluble Hamiltonian because a quantum wave packet that holds together as time increases might be expected to have its classical 
counterparts hold together, for at least small times.

\subsection{Quantum Phase: Approximate Hamiltonian}

For the comparable quantum problem, the Hamiltonian as usual is obtained by replacing in (\ref{22}) the position and momentum variables  $X_{cl}, P_{cl}$  with operators $X,P$:
\begin{equation}\label{25}
H'=\frac{1}{2m}[P-\frac{q}{c}A(x_{cl}(t),t)]^{2}+qV(x_{cl}(t),t)+ (X-x_{cl}(t))qV'(x_{cl}(t),t)-\frac{q}{c}v_{cl}(t)(X-x_{cl}(t))A'(x_{cl}(t),t).
\end{equation}
\noindent Assuming again the wave function form (\ref{13}),
\[
\psi(x,t)=e^{-[x-x_{cl}(t)]^{2}/4\sigma^{2}}e^{i\theta(x,t)}, 
\]
\noindent we obtain the two equations (with neglect of the ``quantum potential" as before): 
\begin{eqnarray}\label{26}
\theta'(x,t)&=&mv_{cl}(t)+\frac{q}{c}A(x_{cl}(t),t),\nonumber\\
-\dot\theta(x,t)&=&\frac{m}{2}v_{cl}^{2}(t)+qV(x_{cl}(t),t)+q(x-x_{cl}(t))V'(x_{cl}(t),t)-\frac{q}{c}v_{cl}(t)(x-x_{cl}(t))A'(x_{cl}(t),t).\nonumber\\
\end{eqnarray}
From the first of Eqs.(\ref{26}),using the second of Eqs.(\ref{10}) again, we have
\begin{eqnarray}\label{27}
\partial_{t}\partial_{x}\theta(x,t)&=&\partial_{t}[mv_{cl}(t)+\frac{q}{c}A(x_{cl}(t),t)]\nonumber\\
&=&-qV'(x_{cl}(t),t)-\frac{q}{c}\dot A(x_{cl}(t),t)+\frac{q}{c}[\dot A(x_{cl}(t),t)+v_{cl}(t)A'(x_{cl}(t),t)]\nonumber\\
&=&-qV'(x_{cl}(t),t)+\frac{q}{c}v_{cl}(t)A'(x_{cl}(t),t).
\end{eqnarray}
\noindent  We see from the second of (\ref{26}) that this is equal to $\partial_{x}\partial_{t}\theta(x,t)$, so $\theta(x,t)$ can be found. 
 
By integrating the first of (\ref{26}),  we find that $\theta(x,t)$ has the form
\begin{equation}\label{28}
\theta(x,t)=x[mv_{cl}(t)+\frac{q}{c}A(x_{cl}(t),t)]+\alpha(t),
\end{equation}
 \noindent where $\alpha(t)$ is an arbitrary function of $t$.  Taking the time derivative of this, and comparing it with the second of (\ref{22}), we find:
\begin{eqnarray}\label{29}
\dot\alpha(t)&=&-\frac{m}{2}v^{2}_{cl}(t)+x_{cl}(t)[qV'(x_{cl}(t),t)-\frac{q}{c}v_{cl}(t)A'(x_{cl}(t),t)] -qV(x_{cl}(t),t),\nonumber\\
&=&-\frac{1}{2m}[p_{cl}(t)-\frac{q}{c}A(x_{cl}(t),t)]^{2}-x_{cl}(t)\frac{d}{dt}p_{cl}(t)-qV(x_{cl}(t),t),\nonumber\\
&=&-\frac{1}{2m}p_{cl}^{2}(t)+\frac{q}{mc}p_{cl}(t)A(x_{cl}(t),t) -\Big[\frac{d}{dt}[x_{cl}(t)p_{cl}(t)]-v_{cl}(t)p_{cl}(t)\Big]-qV(x_{cl}(t),t),\nonumber\\
&=&-\frac{1}{2m}p_{cl}^{2}(t)+\frac{q}{mc}p_{cl}(t)A(x_{cl}(t),t) -\frac{d}{dt}[x_{cl}(t)p_{cl}(t)]+\frac{1}{m}[p_{cl}(t)-\frac{q}{c}A(x_{cl}(t),t)]p_{cl}(t)-qV(x_{cl}(t),t),\nonumber\\
&=&\frac{1}{2m}p_{cl}^{2}(t) -\frac{d}{dt}[x_{cl}(t)p_{cl}(t)]-qV(x_{cl}(t),t).
\end{eqnarray}

Putting the integral of (\ref{29}) into (\ref{28}), we obtain the expression for the phase (up to an additive constant, in which $x_{cl}(0)p_{cl}(0)$ has been absorbed): 
\begin{equation}\label{30}
\theta(x,t)=p_{cl}(t)[x-x_{cl}(t)]+\int_{0}^{t}dt'\frac{1}{2m}p_{cl}^{2}(t')-q\int_{0}^{t}dt'V(x_{cl}(t'),t').
\end{equation}

(As a check, it is shown in Appendix B that when the electric force only depends upon time, this expression for the phase is equal to the phase in Eq. (\ref{8}).)

\section{ Interference: Extra Phase in Two Particle and N+1 Particle Cases.}

We wish to consider an interference situation governed by a Hamiltonian describing many particles interacting with a single particle. 
The situation is such that each particle's wave function 
is expected to be well-approximated as a spatially localized packet with well-defined momentum, and the total wave function is expected to be well-approximated 
as the product of these wave packets or the linear combination of such products. We turn now to show how how to analyze this situation in a consistent fashion.  
Our result  (\ref{13}) for the amplitude and (\ref{30})  for the phase associated with the motion of a single particle in known space-time dependent potentials, 
shall be utilized in this endeavor. 

In the next section we shall consider the simplest problem, an interference situation involving just two particles, which however displays all the features of the more complicated problem.  
 
\subsection{ Two Particle Interaction: Wave Function. } 

The Schro\"dinger equation for two particles moving under their mutual vector and scalar potentials, has the form
\begin{equation}\label{31}
i\frac{d}{dt}|\Psi,t\rangle=\Bigg[\frac{{\bf p}_{1}^{2}}{2m_{1}}+\frac{{\bf p}_{2}^{2}}{2m_{2}}+V(r_{12})-\frac{q_{1}q_{2}}{m_{1}m_{2}c^{2}}{\cal S}[{\bf p}_{1}\cdot {\bf D}^{\leftrightarrow}({\bf r}_{12})\cdot{\bf p}_{2}
]\Bigg]\Psi,t\rangle. 
\end{equation}
\noindent Here, $r_{12}\equiv|{\bf x}_{1}-{\bf x}_{2}|$, ${\bf D}^{\leftrightarrow}({\bf r}_{12})$ is a dyad  and {\cal S}[] is an operation that makes the argument suitably Hermitian.  
In the Lorenz gauge\footnote{Since the  vector potential term is of order $c^{-1}$, one might add the term in the scalar potential to that order.  However, since the Lorenz gauge condition  ${\bf \nabla}\cdot{\bf A}+c^{-1}\dot V=0$ just uses the $0$th order term in the scalar potential, the extra term is not needed for that, nor does it affect anything in our calculation, so it may be omitted} and the Coulomb gauge, $V(r_{12})= q_{1}q_{2}/r_{12}$, while the dyad for these two gauges (the Darwin Hamiltonian in the case of the Coulomb gauge) is, respectively,
\begin{eqnarray}\label{32}
D^{\leftrightarrow}({\bf r}_{12})&=&\frac{{\bf 1}^{\leftrightarrow}}{r_{12}}, \quad
D^{\leftrightarrow}({\bf r}_{12})=
\frac{1}{2}\Big[\frac{{\bf 1}^{\leftrightarrow}}{r_{12}}
+\frac{{\bf r}_{12}{\bf r}_{12}}{r_{12}^{3}}\Big], \hbox{ so the vector potential e.g., due to particle 2 is } \nonumber\\
{\bf A}_{2}&=&\frac{q_{2}}{m_{2}c}{\cal S}\Big[\frac{{\bf p}_{2}}{r_{12}}\Big],  \quad {\bf A}_{2}=\frac{q_{2}}{2m_{2}c}{\cal S}\Big[\frac{{\bf p}_{2}}{r_{12}}+ 
\frac{{\bf r}_{12} {\bf r}_{12}\cdot{\bf p}_{2}}{r_{12}^{3}} \Big] ,
\end{eqnarray}
\noindent  the difference arising from the source of the vector potential being respectively  the current and the transverse component of the current. 

A general solution of  (\ref{31}) will be an entangled state for the two particles. However, for the physical situations we contemplate, 
we suppose the state vector is well-approximated as a direct product of state vectors for each particle. The associated wave functions  
are to be those we have been considering, well localized wave packets moving under 
the vector and scalar potentials due to the other particle, where the other particle's operators are replaced by their classical counterparts. 
How do we make a consistent approximation? 

Consider the variational principle for the Schr\"odinger equation,
\[
\delta\int_{0}^{T}dt\langle\Psi,t|\Big[i\frac{d}{dt}-H\Big]|\Psi,t\rangle=0.
\]
Inserting the approximate solution $|\Psi,t\rangle=|\psi_{1},t\rangle|\psi_{2},t\rangle$, and varying separately for each particle state vector, we obtain for particle 1
(and similarly for particle 2),
\begin{eqnarray}\label{33}
&&i\frac{d}{dt}|\psi_{1},t\rangle +\Big[\langle\psi_{2},t|\Big[ i\frac{d}{dt}- \frac{{\bf p}_{2}^{2}}{2m_{2}}\Big] |\psi_{2},t\rangle \Big]
|\psi_{1},t\rangle\nonumber\\
&&\quad =\Bigg[\frac{{\bf p}_{1}^{2}}{2m_{1}}+
\langle\psi_{2},t|V(r_{12})|\psi_{2},t\rangle-\frac{q_{1}q_{2}}{m_{1}m_{2}c^{2}}\cal S}[{\bf p}_{1}\cdot\langle\psi_{2},t|{{\bf D}^{\leftrightarrow}({\bf r}_{12})\cdot{\bf p}_{2}
|\psi_{2},t\rangle]\Bigg]|\psi_{1},t \rangle \nonumber\\
&&\quad\approx \Bigg[\frac{{\bf p}_{1}^{2}}{2m_{1}}+
V(|{\bf x}_{1}-{\bf x}_{2 cl}(t)|)-\frac{q_{1}q_{2}}{m_{1}m_{2}c^{2}}{\cal S}[{\bf p}_{1}\cdot{\bf D}^{\leftrightarrow}({\bf x}_{1}-{\bf x}_{2 cl}(t))\cdot{\bf p}_{2cl}(t)]
\Bigg]|\psi_{1},t\rangle,
\end{eqnarray}
\noindent where  the approximation in the last step uses the localized nature of packet 2 to replace the operators ${\bf x}_{2 }, {\bf p}_{2 }$ 
by their mean values ${\bf x}_{2cl }(t) ,{\bf p}_{2cl }(t)$. 

Eq.(\ref{33}) is just the equation we have solved for the magnitude and phase  of a localized packet moving in an external vector and scalar potential 
\textit{except} that there is an extra phase term on the left hand side, which we denote
\begin{equation}\label{34}
\dot\phi_{j}(t)\equiv \langle\psi_{j},t|\Big[ i\frac{d}{dt}- \frac{{\bf p}_{j}^{2}}{2m_{j}}\Big]|\psi_{j},t\rangle, \quad j=1,2.
\end{equation}
\noindent (That this is a real number may be seen by taking its complex conjugate and utilizing $d\langle\psi_{j},t|\psi_{j},t\rangle/dt=0.$)

Now, the equations for $|\psi_{1},t\rangle, |\psi_{2},t\rangle$ are form invariant under the replacements  
$|\psi_{1},t\rangle=|\psi_{1},t\rangle'e^{i\beta(t)},  |\psi_{2},t\rangle=|\psi_{2},t\rangle'e^{-i\beta(t)}$: this is to be expected, since the product wave function 
is unchanged by this phase transformation.  This gives the freedom to choose $\beta(t)$ to remove the extra phase term from, say, the equation for $|\psi_{1},t\rangle$.
Given a solution for $|\psi_{1},t\rangle, |\psi_{2},t\rangle$, choose $\beta(t)=\phi_{2}(t)$, so the transformed equations become
\begin{subequations}
\begin{eqnarray}\label{35}
&&i\frac{d}{dt}|\psi_{1},t\rangle' = \Bigg[\frac{{\bf p}_{1}^{2}}{2m_{1}}+
V(|{\bf x}_{1}-{\bf x}_{2 cl}(t)|)-\frac{q_{1}q_{2}}{m_{1}m_{2}c^{2}}{\cal S}[{\bf p}_{1}\cdot{\bf D}^{\leftrightarrow}({\bf x}_{1}-{\bf x}_{2 cl}(t))\cdot{\bf p}_{2cl}(t)]
\Bigg]|\psi_{1},t\rangle',\label{35a}\\
&&i\frac{d}{dt}|\psi_{2},t\rangle' +\Big[\thinspace'\langle\psi_{1},t|\Big[ i\frac{d}{dt}- \frac{{\bf p}_{1}^{2}}{2m_{1}}\Big]|\psi_{1},t\rangle'\Big]|\psi_{2},t\rangle'\nonumber\\
&&\qquad\qquad =\Bigg[\frac{{\bf p}_{2}^{2}}{2m_{2}}+
V(|{\bf x}_{2}-{\bf x}_{1 cl}(t)|)-\frac{q_{1}q_{2}}{m_{1}m_{2}c^{2}}{\cal S}[{\bf p}_{2}\cdot{\bf D}^{\leftrightarrow}({\bf x}_{2}-{\bf x}_{1 cl}(t))\cdot{\bf p}_{1cl}(t)]
\Bigg]|\psi_{2},t\rangle'.\label{35b}\nonumber\\
\end{eqnarray}
\end{subequations}\
\noindent 
By subtracting $\frac{{\bf p}_{1}^{2}}{2m_{1}}|\psi_{1},t\rangle'$ from  (\ref{35a}), and taking the scalar product with $'\langle\psi_{1},t|$, we see that the extra phase in (\ref{35b}) is given by 
\begin{equation}\label{36}
\dot\phi'_{1}(t)\equiv '\negmedspace\negthinspace\langle\psi_{1},t|\Big[ i\frac{d}{dt}- \frac{{\bf p}_{1}^{2}}{2m_{1}}\Big]|\psi_{1},t\rangle'=V(|{\bf x}_{1cl}(t)-{\bf x}_{2 cl}(t)|)-\frac{q_{1}q_{2}}{m_{1}m_{2}c^{2}}{\bf p}_{1cl}(t)\cdot{\bf D}^{\leftrightarrow}({\bf x}_{1cl}(t)-{\bf x}_{2 cl}(t))\cdot{\bf p}_{2cl}(t)
\end{equation}
\noindent That is, \textit{this phase is the time integral of the interaction energy term of the Hamiltonian, where all operators are replaced by their classical values}.

We also note from (\ref{35b}), by subtracting $\frac{{\bf p}_{2}^{2}}{2m_{2}}$ and taking the scalar product with $'\langle\psi_{2},t|$, that 
\begin{equation}\label{37}
\thinspace'\negthinspace\langle\psi_{2},t|\Big[ i\frac{d}{dt}- \frac{{\bf p}_{2}^{2}}{2m_{2}}\Big]|\psi_{2},t\rangle'+'\negmedspace\langle\psi_{1},t|\Big[ i\frac{d}{dt}- \frac{{\bf p}_{1}^{2}}{2m_{1}}\Big]|\psi_{1},t\rangle'= \dot\phi'_{1}(t), \hbox{ that is, }\thinspace'\negthinspace\langle\psi_{2},t|\Big[ i\frac{d}{dt}- \frac{{\bf p}_{2}^{2}}{2m_{2}}\Big]|\psi_{2},t\rangle'=0.
\end{equation}

If we write $|\psi_{2},t\rangle'\equiv e^{i\phi'_{1}(t)}|\psi_{2},t\rangle''$, then $|\psi_{2},t\rangle''$ satisfies (\ref{35b}) without the extra phase term.
Then it, and the solution of (\ref{35a}), are both  
the solutions we have found for 
a localized packet moving in external  potentials, where those potentials now are internal potentials so to speak, potentials due to the other particle 
with operator values replaced by classical ones. So, the wavefunction which solves (\ref{31}) approximately is the product of these two solutions, \textit{multiplied by the 
extra phase factor} $e^{i\phi'_{1}(t)}$.

\subsection{Two Particle Interaction: Interference. }
We shall now show, in an interference experiment involving the two particles interacting, 
that each particle makes the \textit{same} contribution to the phase shift, each the \textit{negative} of the contribution of  (\ref{36}), so the 
\textit{net} phase shift is the negative of the time integral of (\ref{36}). Thus we see the germ here of what will be fleshed out later, 
that the motion of the electron (think particle 1) in the vector potential of the solenoid, and the motion of the solenoid (think particle 2) in the vector potential 
of the electron each make identical contributions to the phase shift. That would give twice the expected phase shift except that \textit{the proper approximation generates an additional  phase shift which is the negative of these}, and so one ends up with the expected phase shift,  the shift due to \textit{either} mechanism.

Consider an interference experiment where, at time 0, packets of the two particles split in two, with equal amplitude both particles going on trajectories labeled $A$, 
or on trajectories labeled $B$. The wave function at some time $t<T$ may be written as
\begin{equation}\label{38}
\Psi(x_{1},x_{2}, t)=\frac{1}{\sqrt{2}}[\psi_{1}^{'A} (x_{1},t)\psi_{2}^{''A} (x_{2},t)e^{i\phi_{1}^{'A}(t)}
+\psi_{1}^{'B} (x_{1},t)\psi_{2}^{''B} (x_{2},t)e^{i\phi_{1}^{'B}(t)}]. 
\end{equation}
\noindent Here, $\psi_{1}^{'}(x_{1},t)=\langle x_{1}|\psi_{1}, t\rangle', \psi_{2}^{''}(x_{1},t)=\langle x_{2}|\psi_{2}, t\rangle''$ satisfy their respective 
Schr\"odinger equations, motion under potentials where the position and momentum operators of the other particle have been replaced by their time-dependent classical values, with no extra phase terms. $\phi_{1}^{'A,B}(t)$ is the extra phase factor for trajectory $A$ or $B$.

The packets finally come together at time  $T$. We suppose that particle 2's packets  
precisely overlap then and thereafter.  We suppose that particle 1's packets meet at (the particle analog of) a half-silvered mirror, coming from opposite directions. If particle 1 had been on path $A$, it splits into two packets that go to the right or left with equal amplitude $1/\sqrt{2}$. 
If particle 1 had been on path $B$, one packet goes  to the right with amplitude $1/\sqrt{2}$ and the other to the left with amplitude $-1/\sqrt{2}$.     The probability of particle 1 being detected either at the left or the right is what is measured.
The wave function at time $T$ (trivially extended thereafter) is therefore  $\Psi(x_{1},x_{2}, T)=\frac{1}{\sqrt{2}}[\Psi_{+}(x_{1},x_{2}, T)+\Psi_{-}(x_{1},x_{2}, T)]$, with 

\begin{equation}\label{39}
\Psi_{\pm}(x_{1},x_{2}, T)=\frac{1}{\sqrt{2}}[\psi_{1}^{'A} (x_{1},T)\psi_{2}^{''A} (x_{2},T)e^{i\phi_{1}^{'A}(T)}
\pm\psi_{1}^{'B} (x_{1},T)\psi_{2}^{''B} (x_{2},T)e^{i\phi_{1}^{'B}(T)}]
\end{equation}
\noindent with the subscript $+$ referring to what is measured to the right and $-$ referring to what is measured to the left, and the superscripts $A,B$ denoting the path.
The individual particle wave functions in (\ref{39}) are of the form (\ref{13}), with phase (\ref{30}), e.g.,
\begin{equation}\label{40}
\psi_{1}^{'A}(x_{1},T)=Ne^{-[x_{1}-x_{1,cl}^{A}(T)]^{2}/4\sigma^{2}}e^{i\big[p_{1,cl}^{A}(T)[x_{1}-x_{1,cl}^{A}(T)]+
\int_{0}^{T}dt\frac{1}{2m}[p_{1,cl}^{A}(t)]^{2}-q_{1}\int_{0}^{T}dtV_{1}(x_{1,cl}^{A}(t),t)\big]}.
\end{equation}

The probabilities of the two outcomes are
\begin{eqnarray}\label{41}
P_{\pm}&\equiv&\int dx_{1}dx_{2}|\Psi_{\pm}
(x_{1},x_{2}, T)|^{2}=\frac{1}{4}\Bigg[2\pm
e^{i(\phi_{1}^{'A}(T)-\phi_{1}^{'B}(T))}\int dx_{1}\psi_{1}^{'A} (x_{1},T)\psi_{1}^{*'B}(x_{1},t)\int dx_{2}\psi_{2}^{''A} (x_{2},T)\psi_{2}^{*''B} (x_{2},t) +cc\nonumber  \Big] \\       
\end{eqnarray}
\noindent Using the expressions (\ref{40}), the integrals in (\ref{41}) are readily performed and give 1, since the packets completely overlap at time $T$ so $x_{cl}^{A}(T)=x_{cl}^{B}(T),p_{cl}^{A}(T)=p_{cl}^{B}(T)$.  We are left then with the phase factors.   
From  (\ref{40}), the  factor associated to interference of each particle is (leaving off the primes):
\begin{eqnarray}\label{42}
e^{i(\Phi_{i}^{A}-\Phi_{i}^{B})}\equiv\int_{-\infty}^{\infty}dx_{i}\psi_{i}^{A}(x_{i},T)\psi_{i}^{*B}(x_{i},T)&=&
   e^{i\int_{0}^{T}dt\frac{1}{2m_{i}}[p_{i,cl}^{A}(t)]^{2}-i\int_{0}^{T}dt\frac{1}{2m_{i}}[p_{i,cl}^{B}(t)]^{2}}e^{-iq_{i}[\int_{0}^{T}dt(V(x_{i,cl}^{A}(t),t)-V(x_{i,cl}^{B}(t),t)]}.\nonumber\\
\end{eqnarray}
Then, the probabilities are
\begin{equation}\label{43}
P_{\pm}=\frac{1}{2}\Big[1\pm\cos[(\Phi_{1}^{A}+\Phi_{2}^{A}+\phi_{1}^{'A})-( \Phi_{1}^{B}+\Phi_{2}^{B}+\phi_{1}^{'B}) ]\Big].
\end{equation}

In order to evaluate the phase in (\ref{42}), we utilize Eq.(\ref{11}). We find, for trajectory $A$ or $B$, 
\begin{subequations}
\begin{eqnarray}\label{44}
p_{i,cl}(t)&=&m_{i}v_{i,0}+\frac{q_{i}}{c}A(x_{i,cl}(t),t)+q_{i}\int_{0}^{t}dt{\cal E}(x_{i,cl}(t),t),\label{44a}\\
x_{i,cl}(t)&=&x_{i,0}+v_{i,0}t+\frac{q_{i}}{m_{i}}\int_{0}^{t}dt'\int_{0}^{t'}dt''{\cal E}(x_{i,cl}(t''),t'').\label{44b}
\end{eqnarray}
\end{subequations}
Since $x_{i,cl}^{A}(T)=x_{i,cl}^{B}(T)$, it follows from (\ref{44b}) that
\begin{equation}\label{45}
\int_{0}^{T}dt\int_{0}^{t}dt'[{\cal E}(x_{i,cl}^{A}(t'),t')-{\cal E}(x_{i,cl}^{B}(t'),t')]=0.
\end{equation}
We may write the momentum dependent phase  for each particle that appears in (\ref{42}), using (\ref{44a}), (\ref{45}), as:
\begin{eqnarray}\label{46}
&&\int_{0}^{T}dt\frac{1}{2m_{i}}[p_{i,cl}^{A}(t)]^{2}-\int_{0}^{T}dt\frac{1}{2m_{i}}[p_{i,cl}^{B}(t)]^{2}=\int_{0}^{T}dt\frac{1}{2m_{i}}
[p_{i,cl}^{A}(t)+p_{i,cl}^{B}(t)][p_{i,cl}^{A}(t)-p_{i,cl}^{B}(t)]\nonumber\\
&&\approx\int_{0}^{T}dtv_{i}(0)\Bigg[\frac{q_{i}}{c}[A(x_{i,cl}^{A}(t),t)-A(x_{i,cl}^{B}(t),t)]+q_{i}\int_{0}^{t}dt'[{\cal E}(x_{i,cl}^{A}(t'),t')-{\cal E}(x_{i,cl}^{B}(t'),t')]\Bigg]\nonumber\\
&&=\int_{0}^{T}dtv_{i}(0)\frac{q_{i}}{c}[A(x_{i,cl}^{A}(t),t)-A(x_{i,cl}^{B}(t),t)]\approx\int_{0}^{T}dt\Big[v_{i, cl}^{A}(t)\frac{q_{i}}{c}A(x_{i,cl}^{A}(t),t)
-v_{i, cl}^{B}(t)\frac{q_{i}}{c}A(x_{i,cl}^{B}(t),t)\Big]\nonumber\\
&&=\int_{0}^{T}dt\Big[{\bf v}_{i, cl}^{A}(t)\cdot\frac{q_{i}}{c}{\bf A}(x_{i,cl}^{A}(t),t)
-{\bf v}_{i, cl}^{B}(t)\cdot\frac{q_{i}}{c}{\bf A}(x_{i,cl}^{B}(t),t)\Big]
\end{eqnarray}
In the approximations made in the second and third lines of (\ref{46}), we have used $v_{i}(0)=v_{i,cl}(t)+0(q)$ from (\ref{11}), and dropped terms in the square of the charges, as has been done throughout this paper.  
In the last line, we have reintroduced vector notation, since the vector potential in our equations has always been the component parallel to the velocity of the particle.  

Thus, from (\ref{42}) and (\ref{46}),we have
\begin{equation}\label{47}
\Phi_{i}^{A}-\Phi_{i}^{B}= \int_{0}^{T}dt\Big[{\bf v}_{i, cl}^{A}(t)\cdot\frac{q_{i}}{c}{\bf A}(x_{i,cl}^{A}(t),t)-q_{i}V(x_{i,cl}^{A}(t),t)\Big]-  
 \int_{0}^{T}dt\Big[{\bf v}_{i, cl}^{B}(t)\cdot\frac{q_{i}}{c}{\bf A}(x_{i,cl}^{B}(t),t)-q_{i}V(x_{i,cl}^{B}(t),t)\Big].  
\end{equation}
 That is,  $\Phi_{i}^{A},\Phi_{i}^{B}$ are each the \textit{time integral of the negative of the Hamiltonian interaction energy with operators replaced by classical variables} 
 for that trajectory ($A$ or $B$).   Due to the symmetry of the interaction under exchange of particles 1 and 2, these are the \textit{same},  independent of $i$.  
 \textit{Since this is the negative of the extra phase} 
 $\phi_{i}^{'A}(T),\phi_{i}^{'B}(T)$, we have for the probabilities (\ref{43}):
\begin{subequations} 
 \begin{eqnarray}
P_{\pm}&=&\frac{1}{2}\Big[1\pm\cos(\Phi^{A}-\Phi^{B})\Big] \hbox{ where }\\\label{48a}
\Phi^{A,B} &=&\int_{0}^{T}dt\Big[{\bf v}_{i, cl}(t)^{A,B}\cdot\frac{q_{i}}{c}{\bf A}(x_{i,cl}^{A,B}(t),t)-q_{i}V(x_{i,cl}^{A,B}(t),t)\Big]\nonumber\\
&=&-\int_{0}^{T}dtH_{\hbox{int}}(x_{1cl}^{A,B}(t), x_{2,cl}^{A,B}(t),p_{1,cl}^{A,B}(t),p_{2,cl}^{A,B}(t)) \label{48b}.
 \end{eqnarray}\\
\end{subequations} 
 
 Thus, we have confirmed the assertion made in the  beginning of this section, that the phase shift is the difference for each path of the time integral 
 of the negative of the interaction Hamiltonian with operators replaced by time-dependent classical variables. 
 
 \subsection{$N$ Particles Interacting With a Single Particle: Wave Function.}
 
We shall extend the result of the previous section since we eventually wish to consider an $N$ particle solenoid interacting with an electron. We shall repeat as closely as possible the steps taken in the discussion of the interaction of two particles.

The Schr\"odinger equation for $N$ identical particles of mass $m$, charge $q$, interacting with a single particle of mass $m_{e}$, charge $e$, under mutual vector and scalar potentials has the form
\begin{equation}\label{49}
i\frac{d}{dt}|\Psi,t\rangle=\Bigg[\frac{{\bf p}_{e}^{2}}{2m_{e}}+\sum_{n=1}^{N}\frac{{\bf p}_{n}^{2}}{2m}+\sum_{n=1}^{N}V(r_{en})-\sum_{n=1}^{N}\frac{eq}{m_{e}mc^{2}}{\cal S}[{\bf p}_{e}\cdot {\bf D}^{\leftrightarrow}({\bf r}_{en})\cdot{\bf p}_{n}
]\Bigg]\Psi,t\rangle. 
\end{equation}
\noindent  (We shall call the single particle the ``electron.")

Upon considering the variational principle for the Schr\"odinger equation with 
 the approximate solution $|\Psi,t\rangle=|\psi_{e},t\rangle\prod_{n=1}^{N}|\psi_{n},t\rangle$, and varying separately for each particle state vector, we obtain 
 \begin{subequations}
\begin{eqnarray}
&&i\frac{d}{dt}|\psi_{e},t\rangle +\Bigg[\sum_{n=1}^{N}\langle\psi_{n},t|\Big[ i\frac{d}{dt}- \frac{{\bf p}_{n}^{2}}{2m}\Big]|\psi_{n},t\rangle \Bigg]|\psi_{e},t\rangle\nonumber\\
&&\quad =\Bigg[\frac{{\bf p}_{e}^{2}}{2m_{e}}+
\sum_{n=1}^{N}\langle\psi_{n},t|V(r_{en})|\psi_{n},t\rangle-\frac{eq}{m_{e}mc^{2}}{\cal S}[{\bf p}_{e}\cdot\sum_{n=1}^{N}\langle\psi_{n},t|{\cal S}[{\bf D}^{\leftrightarrow}({\bf r}_{en})\cdot{\bf p}_{n}]
|\psi_{n},t\rangle\Bigg]|\psi_{e},t \rangle \nonumber\\
&&\quad\approx \Bigg[\frac{{\bf p}_{e}^{2}}{2m_{e}}+
\sum_{n=1}^{N}V(|{\bf x}_{e}-{\bf x}_{n cl}(t)|)-\frac{eq}{m_{e}mc^{2}}{\cal S}[{\bf p}_{e}\cdot\sum_{n=1}^{N}{\bf D}^{\leftrightarrow}({\bf x}_{e}-{\bf x}_{n cl}(t))\cdot{\bf p}_{ncl}(t)]
\Bigg]|\psi_{e},t\rangle,\label{50a}\\
&&i\frac{d}{dt}|\psi_{n},t\rangle +\Bigg[\sum_{n'=1, n'\neq n}^{N}\langle\psi_{n'},t|\Big[ i\frac{d}{dt}- \frac{{\bf p}_{n'}^{2}}{2m}\Big]|\psi_{n'},t\rangle 
 + \langle\psi_{e},t|\Big[ i\frac{d}{dt}- \frac{{\bf p}_{e}^{2}}{2m_{e}}\Big]\psi_{e},t\rangle \nonumber\\
 &&\qquad \qquad -\sum_{n'=1, n'\neq n}^{N}V(|{\bf x}_{ecl}(t)-{\bf x}_{n' cl}(t)|) +
 \frac{eq}{m_{e}mc^{2}}{\bf p}_{ecl}(t)\cdot\sum_{n'=1, n'\neq n}^{N}{\bf D}^{\leftrightarrow}({\bf x}_{ecl}(t)-{\bf x}_{n' cl}(t))\cdot{\bf p}_{n'cl}(t)\Bigg]  |\psi_{n},t\rangle\nonumber\\ 
 &&\quad = \Bigg[\frac{{\bf p}_{n}^{2}}{2m}+\langle\psi_{e},t|V({\bf r}_{en})|\psi_{e},t\rangle-\frac{eq}{m_{e}mc^{2}}{\cal S}[{\bf p}_{n}\cdot\langle\psi_{e},t|{\cal S}[
{\bf D}^{\leftrightarrow}({\bf r}_{en})\cdot{\bf p}_{e}]|\psi_{e},t\rangle
\Bigg]|\psi_{n},t \rangle  \nonumber\\      
&&\quad \approx\Bigg[\frac{{\bf p}_{n}^{2}}{2m}+V(|{\bf x}_{n}-{\bf x}_{ecl}(t)|)-\frac{eq}{m_{e}mc^{2}}{\cal S}[{\bf p}_{n}\cdot[
{\bf D}^{\leftrightarrow}({\bf x}_{n}-{\bf x}_{ecl}(t))\cdot{\bf p}_{ecl}(t)]
\Bigg]|\psi_{n},t \rangle.\label{50b}
\end{eqnarray}
\end{subequations}
\noindent (Note the new terms in the  scalar and vector potentials on the second line of (\ref{50b}) that do not appear in the two particle case.)

We can remove the extra phase terms from the $N$ particles by the phase transformation $|\psi_{n},t \rangle=|\psi_{n},t \rangle' e^{i\alpha_{n}(t)}, 
|\psi_{e},t \rangle=|\psi_{e},t \rangle'e^{-i\sum_{n=1}^{N}\alpha_{n}(t)}$.  Eqs.(\ref{50a}),(\ref{50b}) then become
 \begin{subequations}
\begin{eqnarray}
&&i\frac{d}{dt}|\psi_{e},t\rangle' +\Bigg[\sum_{n=1}^{N}\negthinspace '\langle\psi_{n},t|\Big[ i\frac{d}{dt}- \frac{{\bf p}_{n}^{2}}{2m}\Big]|\psi_{n},t\rangle '\Bigg]|\psi_{e},t\rangle'\nonumber\\
&&\quad\approx \Bigg[\frac{{\bf p}_{e}^{2}}{2m_{e}}+
\sum_{n=1}^{N}V(|{\bf x}_{e}-{\bf x}_{n cl}(t)|)-\frac{eq}{m_{e}mc^{2}}{\cal S}[{\bf p}_{e}\cdot\sum_{n=1}^{N}{\bf D}^{\leftrightarrow}({\bf x}_{e}-{\bf x}_{n cl}(t))\cdot{\bf p}_{ncl}(t)]
\Bigg]|\psi_{e},t\rangle',\label{51a}\\
&&i\frac{d}{dt}|\psi_{n},t\rangle' \approx\Bigg[\frac{{\bf p}_{n}^{2}}{2m_{n}}+V(|{\bf x}_{n}-{\bf x}_{ecl}(t)|)-\frac{eq}{m_{e}mc^{2}}{\cal S}[{\bf p}_{n}\cdot[
{\bf D}^{\leftrightarrow}({\bf x}_{n}-{\bf x}_{ecl}(t))\cdot{\bf p}_{ecl}(t)]
\Bigg]|\psi_{n},t \rangle'.\label{51b}
\end{eqnarray}
\end{subequations}
By subtracting $\frac{{\bf p}_{n}^{2}}{2m}|\psi_{n},t \rangle'$ from  (\ref{51b}), and taking the scalar product with $'\langle\psi_{n},t|$, we see that the extra phase in (\ref{51a}) is given by 
\begin{equation}\label{52}
\dot\phi'_{e}(t)\equiv \sum_{n=1}^{N}\negthinspace'\langle\psi_{n},t|\Big[ i\frac{d}{dt}- \frac{{\bf p}_{n}^{2}}{2m}\Big]|\psi_{n},t\rangle'=\sum_{n=1}^{N}V(|{\bf x}_{ncl}(t)-{\bf x}_{e cl}(t)|)-
\frac{eq}{m_{e}mc^{2}}{\bf p}_{ecl}(t)\cdot      \sum_{n=1}^{N}{\bf D}^{\leftrightarrow}({\bf x}_{ecl}-{\bf x}_{ncl}(t))\cdot{\bf p}_{ncl}(t). 
\end{equation}
\noindent Again, we find that the phase $\phi'_{e}(t)$ is the time integral of the interaction energy term of the Hamiltonian, where all operators are replaced by their classical values.

If we write $|\psi_{e},t\rangle'\equiv e^{i\phi'_{e}(t)}|\psi_{j},t\rangle''$, then $|\psi_{j},t\rangle''$ satisfies (\ref{51a}) without the extra phase term.
 So, the wavefunction which solves (\ref{49}) approximately is the product of $N+1$ wave functions of the type we have considered, for the $N$  particles and the electron, 
\textit{multiplied by the 
extra phase factor} $e^{i\phi'_{e}(t)}$.

\subsection{$N$ Particles Interacting With a Single Particle: Interference. }
We shall now consider a situation where the electron goes on either path $A$ or path $B$, and so the $N$ interacting particles likewise 
move on trajectories $A$ or $B$ determined by their interaction with the electron. The electron packets which traveled by route $A$ and $B$ are brought, at time $T$,  to 
exactly overlap at a half-silvered mirror. The electron is measured as either going right or left, just as was the case for particle 1 in the two particle example, However, for each of the $N$ particles, the  wave packet on trajectory $A$ does not precisely overlap with the wave packet on trajectory $B$ at time T (the overlap is presumed the same thereafter).  So, we shall have to 
consider the effect of this on the measured interference.  

The wave function at time $t<T$ may be written as
\begin{equation}\label{53}
\Psi(x_{e},x_{1}, ...  x_{N}, t)=\frac{1}{\sqrt{2}}\Big[e^{i\phi_{e}^{'A}(t)}\psi_{e}^{''A} (x_{e},t)\prod_{n=1}^{N}\psi_{n}^{'A} (x_{n},t)
+e^{i\phi_{e}^{'B}(t)}\psi_{e}^{''B} (x_{e},t)\prod_{n=1}^{N}\psi_{n}^{'B} (x_{n},t)\Big],
\end{equation}
\noindent where $\psi_{e}^{''}(x_{e},t)=\langle x_{e}|\psi_{e}, t\rangle'',  \psi_{n}^{'}(x_{n},t)=\langle x_{n}|\psi_{n}, t\rangle'$ satisfy their respective 
Schr\"odinger equations, motion under potentials where the position and momentum operators of the other interacting particle have been replaced by their time-dependent classical values.  The wave function at time $T$, just after the electron has passed through the half-silvered mirror,  is  
$\Psi(x_{e},x_{1}, ...  x_{N}, T)=\frac{1}{\sqrt{2}}[\Psi_{+}(x_{e},x_{1}, ...  x_{N}, T)+\Psi_{-}(x_{e},x_{1}, ...  x_{N}, T)]$, with 
\begin{equation}\label{54}
\Psi_{\pm}(x_{e},x_{1}, ...  x_{N}, T)=\frac{1}{\sqrt{2}}[e^{i\phi_{e}^{'A}(t)}\psi_{e}^{''A} (x_{e},T)\prod_{n=1}^{N}\psi_{n}^{'A} (x_{n},T)
\pm e^{i\phi_{e}^{'B}(t)}\psi_{e}^{''B} (x_{e},T)\prod_{n=1}^{N}\psi_{n}^{'B} (x_{n},T)].
\end{equation}
The individual particle wave functions in (\ref{54}) are of the form (\ref{40}). 

The probabilities of the two outcomes are
\begin{eqnarray}\label{55}
P_{\pm}&\equiv&\int dx_{e}dx_{1}|\Psi_{\pm}
(x_{1},x_{2}, T)_{\pm}|^{2}\nonumber\\
&=&\frac{1}{4}\Bigg[2\pm e^{i(\phi_{e}^{'A}(T)-\phi_{e}^{'B}(T))}
\int dx_{e}\psi_{e}^{''A} (x_{e},T)\psi_{e}^{*''B}(x_{T},T )\prod_{n=1}^{N}\int dx_{n}\psi_{n}^{'A} (x_{2},T)\psi_{n}^{*'B} (x_{2},T) +cc \Bigg]      
\end{eqnarray}
\noindent We now must perform the integrals in (\ref{55}) using the expressions (\ref{40}).  The integral over $x_{e}$ is readily performed, since the 
electron wave functions completely overlap at time $T$. The integrals over $x_{n}$  are more complicated, since the wave functions do not overlap.

The  factor associated to interference of each particle in (\ref{55}) (leaving off the subscript and the primes) is 
\begin{eqnarray}\label{56}
\int_{-\infty}^{\infty}dx\psi^{A}(x,T)\psi^{B*}(x,T)&\sim&\int_{-\infty}^{\infty}dx
e^{-\frac{(x-x_{cl}^{A}(T))^{2}}{4\sigma^{2}}}e^{-\frac{(x-x_{cl}^{B}(T))^{2}}{4\sigma^{2}}}e^{ip_{cl}^{A}(T)(x-x_{cl}^{A}(T))}e^{-ip_{cl}^{B}(T)(x-x_{cl}^{B}(T))}\nonumber\\
&&   \cdot e^{i\int_{0}^{T}dt\frac{1}{2m}[p_{cl}^{A}(t)]^{2}-i\int_{0}^{T}dt\frac{1}{2m}[p_{cl}^{B}(t)]^{2}}e^{-iq[\int_{0}^{T}dt(V(x_{cl}^{A}(t),t)-V(x_{cl}^{B}(t),t)]}.
\end{eqnarray}

We may now use:
\begin{eqnarray}\label{57}
\int_{-\infty}^{\infty}dx
e^{-\frac{(x-a)^{2}}{4\sigma^{2}}}e^{-\frac{(x-b)^{2}}{4\sigma^{2}}}e^{ip^{1}(x-a)}e^{-ip^{2}(x-b)}&=&
\int_{-\infty}^{\infty}dx e^{-\frac{x^{2}}{2\sigma^{2}}}e^{\frac{x(a+b)}{2\sigma^{2}}}e^{-\frac{a^{2}}{4\sigma^{2}}} e^{-\frac{b^{2}}{4\sigma^{2}}} e^{i(p^{1}-p^{2})x}e^{-ip^{1}a} e^{ip^{2}b} \nonumber\\
&\sim&e^{\frac{\sigma^{2}}{2}[ \frac{(a+b)}{2\sigma^{2}}+i(p^{1}-p^{2})]^{2} }e^{-\frac{a^{2}}{4\sigma^{2}}} e^{-\frac{b^{2}}{4\sigma^{2}}}e^{-ip^{1}a} e^{ip^{2}b} \nonumber\\
&=&e^{-\frac{(a-b)^{2}}{8\sigma^{2}}} e^{-\frac{\sigma^{2}(p^{1}-p^{2})^{2}}{2}}  e^{i\frac{(p^{1}+p^{2})(b-a)}{2}}. \nonumber\\
\end{eqnarray}
\noindent The factor in ({\ref{56}) is then
\begin{eqnarray}\label{58}
&&\langle\psi^{B},T|\psi^{A},T\rangle\equiv|\langle\psi^{B},T|\psi^{A},T\rangle| e^{i\Phi} \nonumber\\
&=& e^{-\frac{(x_{cl}^{A}(T)-x_{cl}^{B}(T))^{2}}{8\sigma^{2}}}e^{-\frac{\sigma^{2}(p_{cl}^{A}(T)-p_{cl}^{B}(T))^{2}}{2}}e^{i\frac{1}{2}(p_{cl}^{A}(T)+p_{cl}^{B}(T))(x_{cl}^{B}(T)-x_{cl}^{A}(T)) +i\frac{1}{2m}\int_{0}^{T}dt\Big[[p_{cl}^{A}(t)]^{2}-[p_{cl}^{B}(t)]^{2}\Big] -iq\int_{0}^{T}dt[V(x_{cl}^{A}(t),t)-V(x_{cl}^{B}(t),t)]}.\nonumber\\
\end{eqnarray}

For the electron, since $x_{cl}^{A}(T)=x_{cl}^{B}(T), p_{cl}^{A}(T)=p_{cl}^{B}(T)$, the magnitude  in (\ref{58}) is 1. The phase analysis is precisely the same as for either particle in 
 the case of two particles, resulting in the phase difference we shall call $\Phi_{e}^{A}-\Phi_{e}^{B}$. $\Phi_{e}^{A,B }$ is the negative time integral of 
 the interaction Hamiltonian with all particle operators replaced by time-dependent classical variables, for that trajectory ($A$ or $B$).  This cancels the extra phase factor $\phi_{e}^{'A}(T)-\phi_{e}^{'B}(T)$. Therefore, the net phase is that of the $N$ particles. 
 
The phase shift contributed by each particle, according to  (\ref{58}), is 
 \begin{eqnarray}\label{59}
  \Phi_{n}&=&\frac{1}{2}(p_{ncl}^{A}(T)+p_{ncl}^{B}(T))(x_{ncl}^{B}(T)-x_{ncl}^{A}(T)) +\frac{1}{2m}\int_{0}^{T}dt\Big[[p_{ncl}^{A}(t)]^{2}-[p_{ncl}^{B}(t)]^{2}\Big]\nonumber\\
  &&\qquad\qquad\qquad\qquad\qquad \qquad  -q\int_{0}^{T}dt[V(x_{ncl}^{A}(t),t)-V(x_{ncl}^{B}(t),t)].
\end{eqnarray}
\noindent For the electron, the first term vanishes, but here it does not. The analysis is the same as previously, using the 
expressions (\ref{11}) for the classical position and momentum in terms of the potentials, recalling that terms proportional to the square of the charges are disregarded: 
\begin{subequations}
\begin{eqnarray}\label{60}
&&\frac{1}{2}(p_{cl}^{A}(T)+p_{cl}^{B}(T))(x_{cl}^{B}(T)-x_{cl}^{A}(T))\approx\frac{1}{2}[2mv_{0}]\frac{q}{m}\int_{0}^{T}dt\int_{0}^{t}dt'[{\cal E}(x_{cl}^{B}(t'),t')-{\cal E}(x_{cl}^{A}(t'),t')],\nonumber\\
&&\frac{1}{2m}\int_{0}^{T}dt\Big[[p_{cl}^{A}(t)]^{2}-[p_{cl}^{B}(t)]^{2}\Big]=\frac{1}{2m}\int_{0}^{T}dt[p_{cl}^{A}(t)+p_{cl}^{B}(t)][p_{cl}^{A}(t)-p_{cl}^{B}(t)]\label{60a}\\
&\approx&\frac{1}{2m}\int_{0}^{T}dt[2mv_{0}]\Big[\frac{q}{c}[A(x_{cl}^{A}(t),t)-A(x_{cl}^{B}(t),t)]+q\int_{0}^{t}dt'[{\cal E}(x_{cl}^{A}(t'),t')-{\cal E}(x_{cl}^{B}(t'),t')]\Big].\label{60b}
\end{eqnarray}
\end{subequations}
\noindent Adding (\ref{60a}),(\ref{60b}), the double integral of the electric field vanishes, and putting the sum into  (\ref{59})  gives the phase contribution  of the $n$th particle,     
\begin{eqnarray}\label{61}
\Phi_{n}&=&\int_{0}^{T}dt\frac{qv_{n0}}{c}[A(x_{ncl}^{A}(t),t)-A(x_{ncl}^{B}(t),t)]-q\int_{0}^{T}dt[V(x_{ncl}^{A}(t),t)-V(x_{ncl}^{B}(t),t)]\nonumber\\
&\approx&\int_{0}^{T}dt\frac{qv_{ncl}(t)}{c}[A(x_{ncl}^{A}(t),t)-A(x_{ncl}^{B}(t),t)]-q\int_{0}^{T}dt[V(x_{ncl}^{A}(t),t)-V(x_{ncl}^{B}(t),t)]\nonumber\\
&=&\int_{0}^{T}dt\frac{q{\bf v}_{ncl}(t)}{c}\cdot [{\bf A}(x_{ncl}^{A}(t),t)-{\bf A}(x_{ncl}^{B}(t),t)]-q\int_{0}^{T}dt[V(x_{ncl}^{A}(t),t)-V(x_{ncl}^{B}(t),t)]
\end{eqnarray}

Therefore, the phase contributed by all $N$ particles is
\begin{eqnarray}\label{62}
\Phi_{N}&\equiv &\Phi_{N}^{A}-\Phi_{N}^{B}=\sum_{n=1}^{N}\Phi_{n}\nonumber\\
&=&\int_{0}^{T}dt\sum_{n=1}^{N} \Big[\frac{q{\bf v}_{ncl}(t)}{c}\cdot{\bf A}(x_{ncl}^{A}(t),t)-qV(x_{ncl}^{A}(t),t)\Big]
-\int_{0}^{T}dt\sum_{n=1}^{N} \Big[\frac{q{\bf v}_{ncl}(t)}{c}\cdot{\bf A}(x_{ncl}^{B}(t),t)-qV(x_{ncl}^{B}(t),t)\Big].
\end{eqnarray}
Thus, we see that the phase $\Phi_{N}^{A,B} =\Phi_{e}^{A,B}\equiv \Phi^{A,B}$ is the negative of the interaction Hamiltonian with operators replaced
by their classical counterparts for the associated trajectory.  \textit{Again we see the crucial effect of the extra phase}: since  $\Phi^{A}- \Phi^{B}=- [\phi_{e}^{'A}(T)-\phi_{e}^{'B}(T)]$, we have for the probabilities (\ref{55}),
using (\ref{58}),
\begin{equation}\label{63}
P_{\pm}=\frac{1}{2}\Big[1\pm   \prod_{n=1}^{N}e^{-\frac{(x_{ncl}^{A}(T)-x_{ncl}^{B}(T))^{2}}{8\sigma^{2}}}e^{-\frac{\sigma^{2}(p_{ncl}^{A}(T)-p_{ncl}^{B}(T))^{2}}{2}}\cos[\Phi^{A}-\Phi^{B}]\Big].
\end{equation}
\noindent When we apply Eq.(\ref{63}) to the magnetic A-B effect, where the $N$ particles comprise the solenoid, we shall evaluate the magnitude of the 
interference term.  

A few remarks are in order before we close this section.

Of course, the wave function of a quantized solenoid under the influence of the classical field of an electron (as done by Vaidman, and verified by us in the next section) has a phase, since all wave functions have a phase. But, why should that phase be precisely the same  as that acquired  by the quantized electron 
moving in the classical potentials of the solenoid?  We have answered that question here.  At heart, it is due to  the symmetry of the potentials under 
exchange of the electron and solenoid variables.

A concise way to see the equality of these phases  is to note that the current of the electron is ${\bf J}_{e}({\bf x},t)=e{\bf v}_{e}(t)\delta ({\bf x}- {\bf x}_{e}(t))$ 
so the phase contribution of the electron which moved in the classical field of the N particles until time $T$ may be expressed as 
\[
\Phi_{e}=\frac{1}{c}\int_{0}^{T} dt\int d{{\bf x}}{\bf J}_{e}({\bf x},t)\cdot {\bf A}_{N}({\bf x},t),
\]
\noindent where ${\bf A}_{N}({\bf x},t)$ is the vector potential due to the $N$ particles.\footnote{ In the Coulomb gauge, either the current or the transverse current ${\bf J}_{eT}({\bf x},t)$
may be put into this equation: $\int d{{\bf x}}{\bf J}_{e}({\bf x},t)\cdot {\bf A}_{N}({\bf x},t)=\int d{{\bf x}}{\bf J}_{eT}({\bf x},t)\cdot {\bf A}_{N}({\bf x},t)$.  This is because ${\bf J}_{e}({\bf x},t)-{\bf J}_{eT}({\bf x},t)$ is a gradient, and its contribution to the integral vanishes, since ${\bf \nabla}\cdot{\bf A}_{N}({\bf x},t)=0$.}  And, the current of the $N$ particles is 
${\bf J}_{N}({\bf x},t)=q\sum_{n=1}^{N}{\bf v}_{n}(t)\delta ({\bf x}- {\bf x}_{n}(t))$, so their phase acquired while moving in the classical field of the electron is 
\[
\Phi_{N}=\frac{1}{c}\int_{0}^{T} dt\int d{{\bf x}}{\bf J}_{N}({\bf x},t)\cdot {\bf A}_{e}({\bf x},t),
\]
However, using integration by parts, 
\begin{eqnarray}
\Phi_{e}&=&\frac{1}{c}\int_{0}^{T} dt\int d{{\bf x}}[-c\nabla^{2}{\bf A}_{e}({\bf x},t)]\cdot{\bf A}_{N}({\bf x},t)=
\frac{1}{c}\int_{0}^{T} dt\int d{{\bf x}}{\bf A}_{e}({\bf x},t)[-c\nabla^{2}\cdot{\bf A}_{N}({\bf x},t)]\nonumber\\
&=&\frac{1}{c}\int_{0}^{T} dt\int d{{\bf x}}{\bf A}_{e}({\bf x},t)\cdot{\bf J}_{N}({\bf x},t)=\Phi_{N}.\nonumber
\end{eqnarray}

\section{Application to the Magnetic A-B Effect.}

We now apply these considerations to the Aharonov-Bohm magnetic effect, where we model the solenoid as a collection of $N$ particles.  

As remarked in the introduction,  Vaidman demonstrated by a semi-classical calculation  that the phase shift associated to the motion of the solenoid under the vector potential of the electric field of the electron is equal to the usual phase shift associated to the motion of the electron under the vector potential of the solenoid. Using our fully quantum calculation, we have confirmed this in the previous section by a general argument. In this section we shall show this by direct calculation. And, we shall show that 
the magnitude of the interference term in (\ref{63}) is essentially 1 for reasonable values of the parameters in our model. 

\subsection{Model}
We suppose that the electron moves in from infinity to location $(x=0, y=-R, z=0)\equiv(R,\phi'=-\frac{\pi}{2},z=0)$ at time $t=0$, where it is split by a beam splitter into two packets , each packet then circulating around the solenoid in a half circle (in the $z=0$ plane) of radius $R$ with speed $u$ (Fig. \ref{Fig.1}). We choose the discontinuity in angle at $-\frac{\pi}{2}\leftrightarrow\frac{3\pi}{2}$.  The right-side packet goes counterclockwise with angle $\phi'(t)=-\frac{\pi}{2}+\frac{ut}{R}$,
the left-side packet goes clockwise with angle $\phi'(t)=\frac{3\pi}{2}-\frac{ut}{R}$, with the packets meeting again after time $T=\frac{\pi R}{u}$ at  $(R, \pi/2, z=0)$ at a second beam splitter, from which the sum of packets emerges from one side and the difference from the other side.

 \begin{figure}[h]
\includegraphics[width=2.5in]{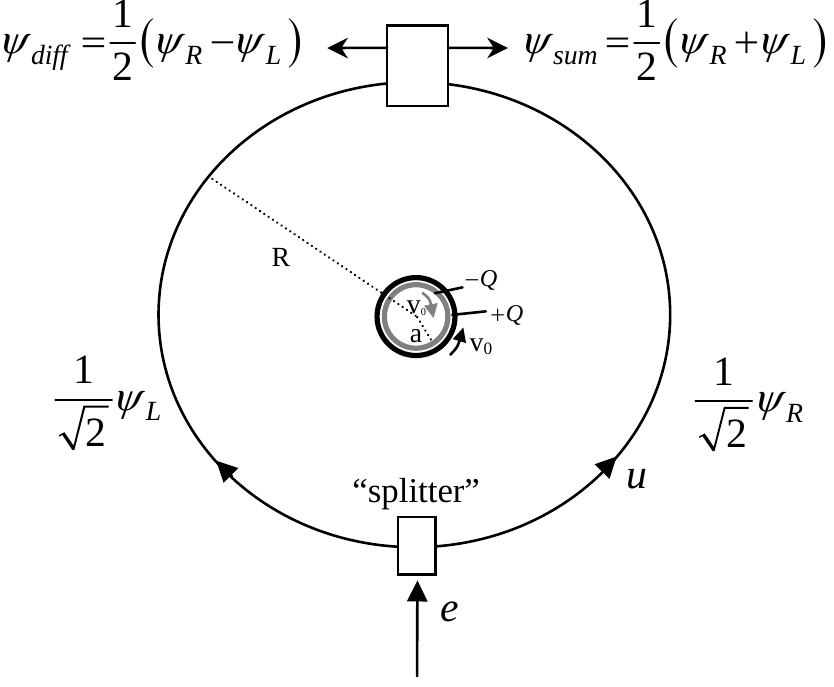}
\caption{The electron packet trajectories in the $z=0$ plane and a cross-section of the rotating cylindrical shells.  The "splitter" boxes do not represent simply
beam splitters but whatever "optics" is necessary to
execute the electron behaviors explained in the text.}
\label{Fig.1}
\end{figure} 

 \begin{figure}[h]
\includegraphics[width=3.5in]{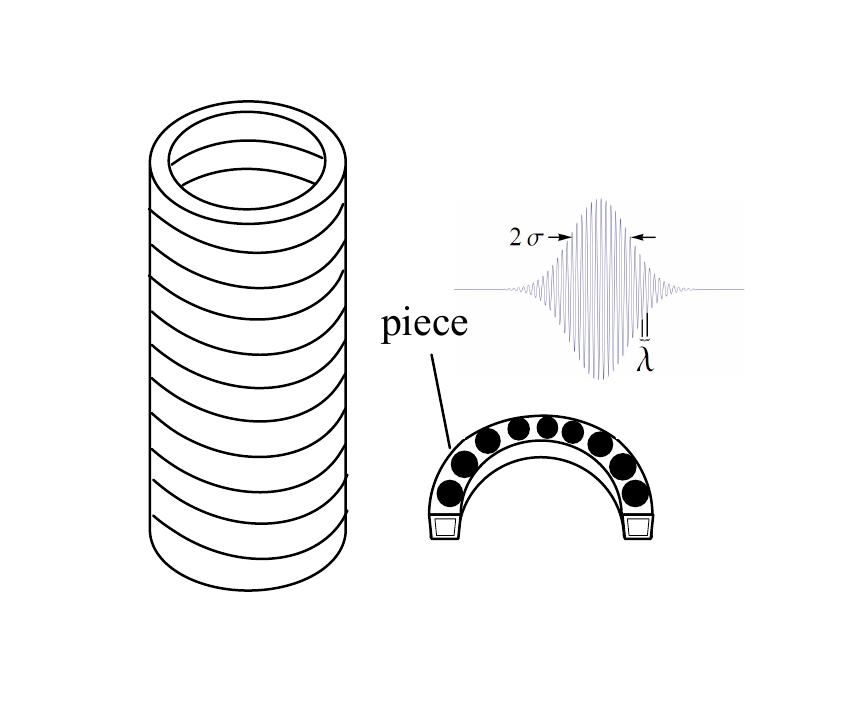}
\caption{  Illustrating one of the two superimposed cylindrical shells comprising the solenoid, a hollow ring within the shell containing charged pieces, and the wave packet of a piece.}
\label{Fig.2}
\end{figure}

 We model the solenoid as comprised of two concentric cylindrical shells of nearly the same radius (which, for simplicity, we shall consider as superimposed). 
Each cylinder  consists of a stack of, say square cross-sectioned, hollow rings (see Fig.\ref{Fig.2}), the center of each ring at $x=0, y=0, z$, with the rings extending from $z=-L/2$ to $z=L/2$.  Each ring is of radius 
$a$,  height characterized by coordinate $z$.  Each ring contains charged pieces, each of mass $m$ and charge $\pm q$, moving with speed $v_{0}$, each of localized extent and well-defined momentum  so that our previous results apply. 
As the electron moves in, its Coulomb field alters the speed of these of these pieces as well as their density, but the current which is their product is 
unchanged\footnote{
Suppose the electron wave packet moves in adiabatically from infinity to $x=0, y=-R, z=0$.  Then,  
 one can neglect the vector potential of the electron as well as neglect $\dot {\bf E}$.  This leaves the electric field as the gradient of the scalar potential. It then follows from Maxwell's equation 
 $\dot {\bf B}=-c{\bf\nabla}\times {\bf E}=0$ that the magnetic field is unchanged. Thus, we see from  $ {\bf J}=(c/4\pi)\bf{\nabla}\times \bf{B}$ that $\dot {\bf J}=0$ so,  the current is unchanged, $J=J_{0}$. Thus, the current associated to the $n$th piece $J_{0}=\sigma_{n}v_{n}$ is unchanged although both surface charge density $\sigma_{n}$ 
 and velocity $v_{n}$ have both been changed from $\sigma_{0}$ and $v_{0}$.  One may further show that $v_{n}(\phi)\approx v_{0}+\frac{qeaR\sin\phi}{mv_{0}D^{3}}.$ 
Thus, we may take the particle densities and velocities as having the constant values  $\sigma_{0}$ and $v_{0}$
since the effect of the correction  term $\sim q$ upon the phase shift is $\sim q^{2}$, which may be ignored.}.

The charge may be written as 
\begin{equation}\label{64}
q=dzd\phi Q/2\pi L,
\end{equation}
\noindent where $\phi$ is the angular coordinate of a piece, $Q$ ($M$) is the magnitude of the total charge (total mass) in one cylindrical shell, and $dz$ is the height of a ring and $ad\phi$ is the length of each piece.  The charge to mass ratio is $q/m=Q/M$ for each piece.

The positively charged pieces move counterclockwise, the negatively charged pieces move clockwise, with uniform surface charge density $\sigma_{0}=Q/2\pi aL$, so there is 
linear current density $J_{0}=Qv_{0}/2\pi aL$  in each ring. Thus,  the magnetic field due to the two cylinders is 
\begin{equation}\label{65}
B_{0}= 2\frac{4\pi}{c}J_{0}=4\frac{v_{0}}{c}\frac{Q}{ a L}        
 \end{equation}
 
 \subsection{Magnitude of Interference Term}
 
One expects the magnitude to differ negligibly from 1.  

For the electron, in either the magnetic or the electric A-B effect,  when the packets recombine at time $T$,  they are brought by the beam splitter to have the same position 
and momentum, 
$x_{ecl}^{1}(T)=x_{ecl}^{2}(T), p_{ecl}^{1}(T)=p_{ecl}^{2}(T)$, which we have seen implies that the electron contributes a factor 1 to the magnitude.

For the electric A-B effect, as discussed in Section VII and Appendix D, the capacitor plate  dynamics is that of a single particle 
(i.e., the plate wave function describes its center of mass behavior) and,  when 
the electron is brought to interfere, the two plates then overlap almost completely in position and momentum, and so their contribution to the magnitude is $\approx 1$. 

The solenoid in the magnetic A-B effect requires a lot more discussion.  We see from (\ref{63}) that there is a dependence upon both position and momentum for each piece. 
Qualitatively, the A-B shift is shared among a large number of pieces (as we shall call the charged particles making up the solenoid). Each piece wave function 
then has such small relative shift (for the electron packet's two traverses) in position and momentum that the magnitude of overlap is $\approx 1$. 
In terms of deBroglie waves, this corresponds to the
need for the deBroglie waves to overlap spatially almost completely, to shift very slightly, and to have nearly the same frequency to get maximal 
interference.

In our model, suppose each ring of radius $a$ contains $n_{a}$ 
pieces of wave packet 
size  $\sigma$ moving clockwise with speed $v_{0}$, each with $n_{e}$ electrons.  Each ring also contains 
$n_{a}$ similar sized pieces  moving counterclockwise with speed $v_{0}$, each with $n_{e}$ ``positive electrons." We note that $\sigma=2\pi a/n_{a}$, and the mass of a piece is 
$m=n_{e}m_{e}$, where $m_{e}$ is the mass of an electron.  The solenoid is of length $L$ and 
contains $n_{L}\equiv L/\sigma$ rings.  The number of pieces is $N=2n_{L}n_{a}$.  

We shall give here an example, to illustrate how, for reasonable values of the parameters, the sum of all the exponents in (\ref{63}) corresponding to all $N$ pieces in the solenoid can be negligibly small.

Say  $a$ is of order 1cm,  $R$ is of order 10 cm, $L$ of order 100cm,  $v_{0}$ is of order 1cm/s (a typical drift velocity of electrons in a conductor), $u$ is of order 1m/sec. 
 
 In terms of the total number of positive or negative electron charges,  $N_{e}=n_{e}n_{a}n_{L}$, the A-B shift is $\Phi_{AB}=4\pi N_{e}\frac{e^{2}}{\hbar c}\frac{v_{0}}{c}\frac{a}{L}$ (see Eq.(\ref{67}), below).  Say the setup gives a shift $\Phi_{AB}=\pi$.  Then, $N_{e}\approx 10^{14}$ electrons.
 
 The strongest constraint on $n_{a}$ is the condition for our model that $\lambda\equiv h/mv_{0}<<\sigma$, i.e., that a solenoid piece wave packet contain many wavelengths so that its momentum is well defined.  Using $m=n_{e}m_{e}=\frac{N_{e}}{n_{a}n_{L}}m_{e}=\frac{N_{e}}{n_{a}}\frac{L}{\sigma}m_{e}$ and $\sigma=2\pi a/n_{a}$,       this constraint becomes 
$n_{a}^{3}<<N_{e}\frac{m_{e}v_{0}a}{\hbar}\frac{2\pi a}{L}\approx 6\times10^{12}$, which we satisfy by choosing $n_{a}\approx 1000$ pieces per ring. 

This determines $m$ and so makes $\lambda\approx 10^{-6}$cm and determines $\sigma\approx 6\times 10^{-3}$cm, so there are $\approx 6000$ wavelengths in a wavepacket. Also, then there are  $n_{e}\approx 6\times 10^{6}$ electrons per piece, $n_{L}\approx1.5\times 10^{4}$ rings in the solenoid and $n_{p}=1.5\times 10^{7}$ pieces in the solenoid.  

For this estimate, we shall treat each piece of the solenoid as participating equally in the phase shift, even though the pieces in the ring in the plane of the electron's motion, being closer to the electron, can receive a larger force and therefore displacement and, also, pieces in the same ring are displaced differently. 

Since the A-B shift  $\Phi_{AB}=\pi$ corresponds to a total shift of all pieces by a distance $\lambda/2\approx 5\times 10^{-7}$cm, each 
individual piece  shifts by a distance   $\lambda/2n_{p}\approx 3\times 10^{-14}$cm.  The squared fractional displacement 
appearing in  the exponent of the first term in (\ref{63}) for one piece is therefore $\frac{(x_{cl}^{1}(T)-x_{cl}^{2}(T))^{2}}{8\sigma^{2}}\approx 3\times 10^{-24}$, and for all $n_{p}=1.5\times 10^{7}$ pieces the spatial displacement total exponent is $\approx 10^{-16}$.

In order to achieve the displacement of $\approx 3\times 10^{-14}$cm in the time $T=\pi R/u\approx  0.3$sec it takes the electron packets to complete their traverse, 
the relative speed change of the pieces is $\Delta v \approx 10^{-13}$cm/s.  Therefore, since $\sigma(p_{cl}^{1}(T)-p_{cl}^{2})(T))/\hbar\approx \sigma m \Delta v/\hbar$, using  
$m\approx n_{e}m_{e}$, we have $\sigma(p_{cl}^{1}(T)-p_{cl}^{2})(T))/\hbar\approx \times 10^{-8}$.  Thus for one piece, the momentum contribution to the 
exponent in 
 (\ref{63}) is therefore $\approx 0.5\times 10^{-16}$, and for all $n_{p}$ pieces the momentum-dependent total exponent is $\approx 10^{-9}$.
 
 We conclude that the magnitude in (\ref{63}) is 1 to high accuracy.

\subsection{Direct Calculation of Phase Shift}

 \begin{figure}[t]
\includegraphics[width=3.5in]{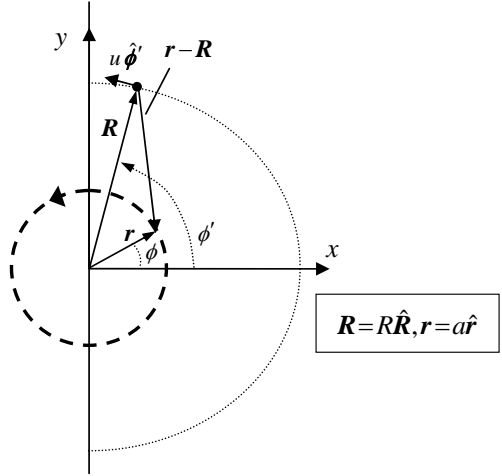}
\caption{Illustration of the notation.}
\label{Fig.3}
\end{figure}

The phase shift of the electron moving in the vector potential of the solenoid, the well-known calculation of the A-B phase shift, is
\begin{equation}\label{66}
\Phi_{AB}=\Phi_{e}^{A}-\Phi_{e}^{B}=\frac{1}{c}\int_{0}^{T} dt eu A(R)-\frac{1}{c}\int_{0}^{T} dt e(-u)(- A(R))=2T \frac{e}{c}u \frac{a^{2}B_{0}}{2R}= \frac{e}{c}\pi a^{2}B_{0}.  ,
\end{equation}
\noindent where $A(R)=\frac{a^{2}B_{0}}{2R}$ is magnitude of the solenoid's vector potential which is parallel to the electron velocity for trajectory $A$ and 
and  antiparallel for trajectory $B$, and we have used $uT=\pi R$. (In the previous section's discussion of the magnitude of the interference term,  we have used the expression
\begin{equation}\label{67}
\Phi_{AB}= \frac{e}{c}\pi a^{2}B_{0}=\frac{e}{c}\pi a^{2}\Big[4\frac{v_{0}}{c}\frac{N_{e}e}{ a L} \Big]\frac{1}{\hbar},
\end{equation}
\noindent where $B_{0}$ comes from (\ref{65}), $Q=N_{e}e$ and the correct factor of $\hbar$ has been inserted. )

Now we explicitly calculate the phase contribution of the pieces of the solenoid due to the field of the electron.  

We shall do this in both the Lorenz gauge 
and the Coulomb gauge. 
At time $t$, the electron is at the location $(R,\phi'(t),0)$, where $\phi'(t)=-\pi/2+ut/R$, and the solenoid piece is at $(a,\phi_{n}(t),z_{n})$ 
where $\phi_{n}(t)=\phi_{n}(0)+v_{0}t/a+o(q)$.  We identify each piece, formerly labeled by the index $n$, by its initial angular value $\phi_{n}(0)$ and its 
 $z=z_{n}$ value.  According to (\ref{32}), the vector potential caused by the electron at the 
location of the piece in the solenoid at time $t$ is 
\begin{subequations}  
\begin{eqnarray}
{\bf A}(\phi_{n}(0), z_{n},t)&=&\frac{e}{ c}\frac{{\bf u}(t)}{|{\bf r}-{\bf R}|}, \label{68a}\\
{\bf A}(\phi_{n}(0), z_{n},t)&=&\frac{e}{ 2c}\frac{{\bf u}(t)}{|{\bf r}-{\bf R}|}+
\frac{e}{ 2c}[{\bf r}-{\bf R}]\frac{{\bf u}(t)\cdot[{\bf r}-{\bf R}]}{|{\bf r}-{\bf R}|^{3}}, \label{68b}
\end{eqnarray}
\end{subequations} 
\noindent in the Lorenz and Coulomb gauges respectively. Here, ${\bf r}={\bf r}(a,\phi_{n}(t),z_{n}), {\bf R}={\bf R}(R,\phi'(t),0)$.

First we treat the Lorenz gauge. We consider the counter-clockwise traverse of the electron (path A) and the phase it gives to the positively charged pieces of the solenoid. 
The contribution to the phase for the $n$th piece is 
\begin{eqnarray}\label{69}
\Phi_{n}^{A}(z_{n},\phi_{n}(0), T)&=&\int_{0}^{T}dt \frac{q}{c}v_{0}\hat \phi(t)\cdot{\bf A}(z,t)\nonumber\\
&=&   \frac{v_{0}equ}{c^{2}}\int_{0}^{T}dt \frac{\cos(\phi_{0}+\frac{v_{0}t}{a} +\pi/2-\frac{ut}{R})}{\sqrt{R^{2}+z^{2}+a^{2}-2aR\cos(\phi_{0}+\frac{v_{0}t}{a} +\pi/2-\frac{ut}{R})}}            
\end{eqnarray}
\noindent where we have used $\hat\phi(t)\cdot{\bf u}(t)=u\hat\phi(t)\cdot\hat\phi'(t)=u\cos(\phi(t)-\phi'(t))$ and  ${\bf r}\cdot {\bf R}=aR\cos(\phi(t)-\phi'(t))$.
We now use   (\ref{64}) to write $q=dzd\phi_{0}Q/2\pi L$ and sum over all pieces, i.e., integrate over $\phi_{0}$ and $z$.  Since   $\phi_{0}$ 
is integrated over a $2\pi$ range,  the argument of the cosine can be replaced by an angle $\theta$ integrated over that range.  Then, we assume that 
the solenoid length $L>>a, R$, which simplifies the integral over $z$
\begin{eqnarray}\label{70}
\Phi_{sol}^{A}&=& \frac{v_{0}Qeu}{2\pi Lc^{2}}\int_{0}^{T}dt\int_{0}^{2\pi}d\theta\int_{-L/2}^{L/2}dz
\frac{\cos\theta}{\sqrt{R^{2}+z^{2}+a^{2}-2aR\cos\theta}}  \nonumber\\ 
&\approx& \frac{v_{0}QeuT}{2\pi Lc^{2}}\int_{0}^{2\pi}d\theta \cos\theta  \Big[2 \ln L-\ln[R^{2}+a^{2}-2aR\cos\theta]\Big]    \hbox{ and, integrating by parts,}
 \nonumber\\  
  &=& \frac{v_{0}QeuT}{2\pi Lc^{2}}\int_{0}^{2\pi}\sin\theta  d\ln[R^{2}+a^{2}-2aR\cos\theta] = \frac{v_{0}QeuT}{2\pi Lc^{2}}2aR\int_{0}^{2\pi}d\theta \frac{\sin^{2}\theta}{R^{2}+a^{2}-2aR\cos\theta}               \nonumber\\
  &=& \frac{v_{0}QeuT}{2\pi Lc^{2}}2aR\frac{1}{R^{2}}\int_{0}^{2\pi}d\theta \frac{\sin^{2}\theta}{1+(a/R)^{2}-2a/R\cos\theta}\nonumber\\
  &=&\frac{v_{0}QeuT}{2\pi Lc^{2}}2aR\frac{1}{R^{2}}\pi=\frac{e}{4c}\pi a^{2}B_{0}
\end{eqnarray}
\noindent where the last integral over $\theta$ is $\pi$ if $a/R<1$, which it is, and we have used $uT=\pi R$ and the expression (\ref{65}) for $B_{0}$.

The result (\ref{70}) is $1/4 $ the A-B shift (\ref{66}). The negatively charged rings give the same result ($Q\rightarrow-Q, v_{0}\rightarrow-v_{0}$), so traverse A 
gives  $1/2 $ the A-B shift. The counterclockwise traverse also gives  $1/2 $ the A-B shift ($u\rightarrow -u$, and the $B$ traverse phase is subtracted).  Thus
we get the full A-B shift.  

For the Coulomb gauge, the first term in the vector potential expression (\ref{68b}) is 1/2 the Lorenz gauge potential (\ref{68a}), so this term gives 1/8 the A-B shift. 
 Using  $\hat\phi(t)\cdot[{\bf r}-{\bf R}]=- \hat\phi(t)\cdot{\bf R}=R \sin(\phi(t)-\phi'(t))$  and ${\bf u}(t)\cdot[{\bf r}-{\bf R}]= {\bf u}(t)\cdot{\bf r}=ua \sin(\phi(t)-\phi'(t))$, the second term in (\ref{68b}) gives the shift, for large $L$:
\begin{eqnarray}\label{71}
\Phi_{sol}^{A}&=&\frac{1}{2} \frac{v_{0}QeuaR}{2\pi Lc^{2}}\int_{0}^{T}dt\int_{0}^{2\pi}d\theta\int_{-L/2}^{L/2}dz
\frac{\sin^{2}\theta}{[R^{2}+z^{2}+a^{2}-2aR\cos\theta]^{3/2}}  \nonumber\\ 
&\approx& \frac{v_{0}QeuTaR}{4\pi Lc^{2}}\int_{0}^{2\pi}d\theta \frac{2\sin^{2}\theta}{R^{2}+a^{2}-2aR\cos\theta}    
 \end{eqnarray}
\noindent This is 1/2 the result given by the last term on the third line in Eq.(\ref{70}), i.e., it is 1/8 of the A-B shift, so both vector potential terms combined in the Coulomb gauge give 
1/4 the A-B phase shift.

\subsection{A time-average approach to understanding the contribution of a solenoid piece to the phase shift.}

Here is a simple, intuitively appealing, approach which gives the result (\ref{70}).

Suppose each piece of the solenoid moves slowly compared to the speed of the electron over the interval $T$.   Then, \  one might consider the approximation of replacing the vector potential 
created by the electron  at the site of a piece, during the piece's traverse along its own path, by the time-averaged potential it experiences.  

This averaging technique will also apply even if the speed of a solenoid pieces is not slow compared to the electron speed, as in the prior section. Because one piece of the solenoid replaces  another during their motion, a piece will always be present at a given place in the solenoid and thus that place can be considered to experience an average vector potential and we can add the contribution from all the places.
This approach starts with the following Hamiltonian for a piece:
\begin{equation}\label{72}
H=\frac{1}{2m}[\hat p-\frac{q}{c}\langle A\rangle_{T}]^{2}\hbox{ where } \langle A\rangle_{T}\equiv\frac{1}{T}\int_{0}^{T}dtA(t).
\end{equation}
\noindent where $\langle A\rangle_{T}\equiv\frac{1}{T}\int_{0}^{T}dtA(t)$ and $A(t)$ is the component of the electron's vector potential along the path of the solenoid piece. 
Because $\langle A\rangle_{T}$ is a constant, it immediately follows (as is usual when treating the phase contributed by the electron's motion) that the phase contributed by the piece is the path integral of the vector potential:
\begin{equation}\label{73}
\hbox{Phase}(T)=\int_{path}ds\frac{q}{c}\langle A\rangle_{T}=v_{0}T\frac{q}{c}\langle A\rangle_{T}.
\end{equation}
\noindent  But, this is precisely the expression given in (\ref{40}) for the phase contributed by the motion of a piece which, when summed over all positively charged pieces as in the previous section, gives 1/4 the A-B phase.
Including, in a similar way,  the  contribution of the negatively charged pieces and the contribution of the left traverse, gives the full A-B shift.  

Thus, the phase shift  can be viewed as due to the action of the \textit{time-averaged} electric-field-producing vector potential 
of the electron acting on the charged pieces of the solenoid as they travel their short paths, the small arcs of the ring in which they are constrained to move.

A more detailed analysis from this point of view appears in Appendix C.

\subsection{Electric Force in the Magnetic A-B Effect.}
We have shown that the standard A-B calculation and the solenoid-involved calculation are two alternative ways of calculating the same thing. 
It is remarkable how conceptually dissimilar they are.  In the usual case, there is the situation of 
a non-field-producing potential causing the phase shift.  We would like to point out in this subsection, in the other case, although we have expressed the phase shift in terms of the vector potential, that it can be 
looked at as due to the electric force of the electron on the solenoid pieces, the view that Vaidman took in his semi-classical calculation.\cite{Vaidman}. 

It may be noted that Vaidman reasoned the
existence of the electric force on the solenoid using
the Faraday law. Faraday's law encompasses two
distinct effects (see e.g. reference 8), one dealing
with the EMF generated in a conductor traveling
through a magnetic field, the other dealing with the
EMF generated when there is a changing magnetic
field inside a closed path. Vaidman's analysis uses
the latter.  He reasoned that, when the electron
accelerates from rest and begins its traverse, it
produces a changing magnetic flux in the solenoid.
Thus, he obtained the electric field induced in the
solenoid (since its integral over the solenoid
circumference is the EMF), whose force changes the
speed of the charged solenoid cylinders. This expression of the second Faraday law effect is equivalent to noting that the accelerating electron causes a time-changing vector potential, whose (negative) time derivative is this
same electric field.

Consider for example the effect of the electron's vector potential, on the positively charged solenoid pieces, during its right-traverse. We shall suppose that the electron starts from rest 
 so that its initial vector potential is 0 (we ignore the solenoid's self-vector potential). We take the time dependence of the electron's vector potential at the site of each solenoid piece 
to be just due to the electron's motion and not the piece's motion, i.e., ${\bf A}({\bf r}_{n},t)$ at the site ${\bf r}_{n}$. This is 
because the replacement of solenoid pieces by successive pieces means we need only label the  pieces' locations and ignore their translation.  Then, the phase associated to these solenoid pieces may be written as
\begin{eqnarray}\label{74}
\Phi^{A}&=&\int_{0}^{T}dt \sum_{n}\frac{q}{c}{\bf v}_{n}\cdot {\bf A}({\bf r}_{n},t)\nonumber\\
&=&-\sum_{n}\int_{0}^{T}d{\bf x}_{n}(t) \cdot\int_{0}^{t}dt'\Big[- \frac{\partial}{\partial t'} \frac{q}{c}{\bf A}({\bf r}_{n},t')\Big]=-\sum_{n}\int_{0}^{T}d{\bf x}_{n}(t) \cdot \int_{0}^{t}dt'{\bf F}_{n}({\bf r}_{n},t').                         
\end{eqnarray}
\noindent We have written ${\bf v}_{n}dt=d{\bf x}_{n}$ as the displacement of the $n$th piece during $dt$. We have noted that the bracketed expression is the electric force ${\bf F}_{n}({\bf r}_{i},t')$ 
exerted by the electron on the $n$th particle 
in the interval $dt'$.  

Thus, this phase shift contribution has been expressed in terms of $dt'{\bf F}_{n}({\bf r}_{n},t')=dp_{n}(t')$, the impulse exerted during $dt'$ by the electron on the solenoid particles. 

If we consider that the electron rapidly accelerates to speed $u$ at the beginning of its trajectory, since the vector potential is proportional to $u(t)$, 
there is a large initial impulse (followed,  as the electron continues its traverse, by a force on each solenoid piece which averages out over all pieces to 0), 
causing a sudden initial change of the momentum $\Delta {\bf p}_{n}$ of the positively charged pieces (and the opposite change in momentum for the negatively charged pieces), so
$\Phi^{A}=-\sum_{n}\int_{0}^{T}d{\bf x}_{n}(t) \cdot\Delta {\bf p}_{n}$.

This impulse-induced differential of momentum, occurring oppositely for the electron's left-traverse, causes the phase shift to accumulate over the interval $T$, which was the point of view taken by Vaidman in his semi-classical calculation. We can see that here by writing $\Delta p_{n}=mv_{n}-mv_{0}$, which is constant and close to parallel to 
the piece displacement $\int_{0}^{T}d{\bf x}_{n}\equiv \Delta \sigma_{n}$ over time $T$. Tossing out the term $mv_{0}$ which does not contribute to the A-B shift as it is the same for left and right traverses of the electron, we have  for the effective $\Phi^{A}=-\sum_{n}\Delta \sigma_{n}mv_{n}=\sum_{n}\Delta \sigma_{n}/\lambda_{n}$, where $\lambda_{n}$ is the deBroglie 
wavelength for the piece.  

\section{Application to the Electric A-B Effect.}
 \begin{figure}[b]
\includegraphics[scale=.8]{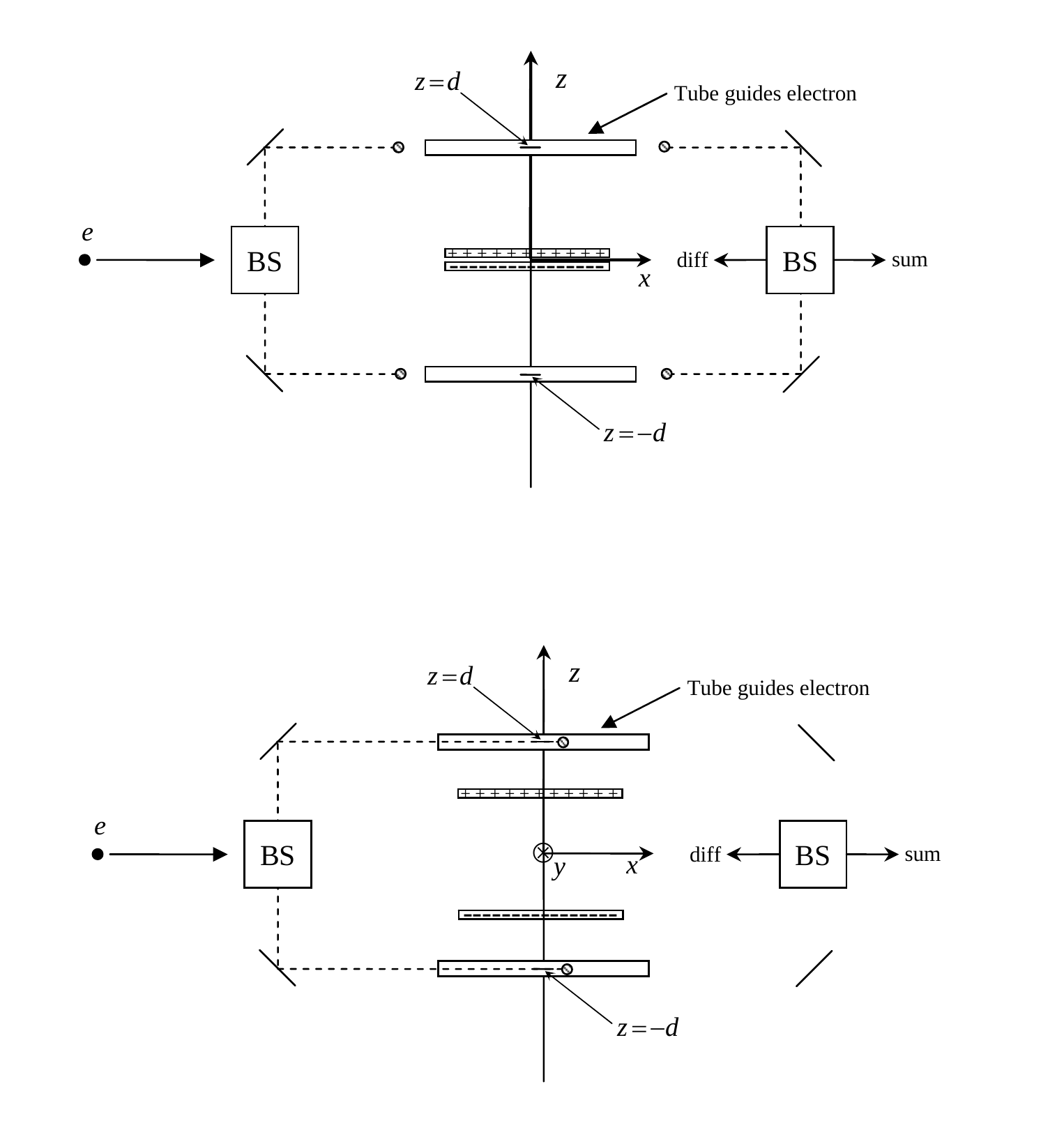}
\caption{ Top: Setup for electric A-B effect, when the plates are together, before and after the electron nears the center of the plates.  Bottom:  
Illustrating the plates with separation $\approx D$ while the electron moves  for a time interval T. (Figures are not to scale, the plates and tubes should be greatly stretched  in the $\pm x$-direction.)}
\label{Fig.4}
\end{figure}  

Lastly, we turn to discuss the electric A-B effect.  This has precisely the same dichotomy as the magnetic A-B effect, two different ways of viewing and calculating the 
A-B phase shift.  However, here it can be shown more simply (because the Coulomb interaction is a direct particle interaction, not an interaction mediated by an intermediate field as is 
the interaction through the vector potential) precisely how this involves two alternative views of the identically same calculation. 

Using a simplified model, where the 
electron sits in a zero-field environment halfway between two identical point charges, again by a semi-classical argument, Vaidman\cite{Vaidman} showed that, while the A-B shift is obtainable as due to the electron's motion in the field of the charges,  it can also be obtained as due to the motion of the charges in the electric field of the electron. 

Instead of Vaidman's simplified model, we shall apply  our considerations to the usually discussed electric A-B effect setup,  
where electron packets passing over and under a charged capacitor are brought together to interfere.   According to the usual view, the phase shift 
$\Phi_{AB}=-e\sigma DT$ (e is the charge on the electron, 
$\sigma$ is the surface charge on the plates, $D$ is the plate separation and $T$ is the time of traverse of the electron---in this section we use rationalized MKS units) is due to the quantized electron moving in the constant potential of the classical capacitor.  We shall show that the same phase shift is obtained by consideration of the quantized capacitor plates moving in the electric field of the classical electron.

However, we do more. We consider what happens when both electron and plates are quantized.  The Hamiltonian is separable, the wave function is the product of wave functions for the electron and plates, and we show how the phase shift of this product wave function, the A-B electric phase shift, can be viewed as either due to the electron motion or due to 
the capacitor plate motion, alternative ways of calculating the same thing.

\subsection{Model}

  We consider two very large plates containing charges ``glued down," with uniform charge densities $\pm\sigma$ and area $A$, lying initially on top of one another in the $x-y$ plane at $z=0$ (see Fig.(\ref{Fig.4})).  

There are two long hollow frictionless tubes parallel to the 
$x$-axis, one at $y=0, z=d$, the other at $y=0, z=-d$, with the entrance and exit of each tube at the plate edges. From the left, an electron approaches  a beam spitter 
which creates two wave packets of equal amplitude., following which 
 two mirrors direct the packets into the tubes.  

After the packets have travelled well away from the capacitor edges, the two plates are separated, rapidly moved by an external field (over a time interval much shorter than $T$)  so that the positive and negative plates 
are left at rest at $\pm D/2$ at time $t=0$.  The plates are free to move thereafter for a time interval $T$, but it is assumed that they are so massive that their displacement is relatively small compared to $D$ over that interval.  

At time $T$, while  the electron is still far from the capacitor edge, the plates are rapidly  returned to the plane $z\approx 0$ by an external  field. (While the plates acquire a phase shift during their separation and rejoining motions,  we  may take the time scale to be brief enough that  it may be neglected compared to the phase shift acquired during the interval $T$.) Eventually the electron reaches the edge of the plates and, by means of two more mirrors reflecting the packets toward each other and another beam splitter, 
the packets emerge with their sum in one direction and their difference in another direction, each direction containing a detector.

The  Hamiltonian over the time interval $(0, T)$ for electron passage above ($H_{+}$) or below ($H_{-}$) the capacitor is
\begin{equation}\label{75}
H_{\pm}=\frac{P_{e}^{2}}{2m}+\frac{P_{U}^{2}}{2M}+\frac{P_{L}^{2}}{2M}+\frac{1}{2}[\sigma^{2}A\pm e\sigma](Z_{U}-Z_{L}).
\end{equation}
\noindent  Here $(X_{e}, P_{e})$ ,  $(Z_{U}, P_{U})$ and $(Z_{L}, P_{L})$ are the conjugate operators for the electron motion in the $x$-direction, upper plate and lower plate motion in the $z-$direction respectively, and $A$ is the plate area. This gives the force on each capacitor plate as due to the other plate and the electron and, of course, no force on the electron.
\subsection{Calculations.}

One can take the point of view that the capacitor is a classical object, and only the electron should be quantized.  Then, one may truncate this Hamiltonian by throwing away the plate 
potential energy and kinetic energy operator terms  and setting $Z_{U}=D/2, Z_{L}=-D/2$, their classical positions to excellent accuracy.  The solution of 
Schr\"odinger's equation with the Hamiltonian
\begin{equation}\label{76}
H_{\pm electron}=\frac{P_{e}^{2}}{2m}+\frac{1}{2}[\pm e\sigma]D
\end{equation}
\noindent is easily seen to have the form 
\begin{equation}\label{77}
\psi_{\pm}(x,T)=\psi'(x,T)e^{ -i\frac{1}{2}[\pm e\sigma]DT},
\end{equation}
\noindent where $\psi'(x,T)$ is a free-particle wave function (this is, of course, what is obtained from Eq.(\ref{8}) or Eqs.(\ref{13}),(\ref{30}) applied to this problem).  $\psi_{+}(x,T)$ ($\psi_{-}(x,T)$ ) 
is the wave function for the electron packet which passes above (below) the capacitor plates.  By inspection, the electric A-B phase shift, the difference of phases for the two traverses,
is seen to be $-e\sigma DT$

 However, one may rewrite the Hamiltonian  (\ref{75}) as (omitting the self-energy of the capacitor since it makes the same contribution to the phase for the upper and lower electron traverses), 
\begin{equation}\label{78}
H_{\pm}=\frac{P_{e}^{2}}{2m}+\Bigg[\frac{P_{U}^{2}}{2M}\pm \frac{1}{2}e\sigma Z_{U}\Bigg]    +     \Bigg[\frac{P_{L}^{2}}{2M}\mp \frac{1}{2}e\sigma Z_{L}\Bigg]. 
\end{equation}
In this form,  the wave function is the product of the free electron wave function, associated to no phase shift, and the upper and lower plate wave functions which 
do have associated phase shifts.  We use our solution Eq.(\ref{8}) in Section II (or Eqs.(\ref{13}),(\ref{30}) in Section III) of  the problem of finding the wave function for one-dimensional motion in an electric potential to obtain the wave function in terms of the classical variables:
\begin{equation}\label{79}
\psi_{\pm}(x, z_{U}, z_{L}, T)=\psi_{el}(x)Ne^{-\frac{(z_{U}-D/2)^{2}}{4\sigma'^{2}}}e^{\mp i \frac{1}{2}e\sigma T[z_{U}-D/2]}e^{\mp i\frac{1}{4}eD\sigma T}\cdot 
e^{-\frac{(z_{L}+D/2)^{2}}{4\sigma'^{2}}}e^{\pm i \frac{1}{2}e\sigma T[z_{L}+D/2]}e^{\mp i\frac{1}{4}eD\sigma T}.   
\end{equation}
\noindent In obtaining (\ref{79})  we have used  $p_{U, cl}(T)=-\frac{1}{2}e\sigma T, p_{L, cl}(T)=-\frac{1}{2}e\sigma T$ (the upper and lower plates move toward each other during T),  $z_{U, cl}(T)\approx D/2, z_{L, cl}(T)\approx -D/2$, and e.g., for the upper traverse situation, $-\int_{0}^{T}dt'V(z_{U}(t'))=-\int_{0}^{T}dt'\frac{e\sigma}{2}\frac{D}{2}=-\frac{1}{4}eD\sigma T$. 

Because of the narrowness of the gaussian wave packets in (\ref{79}), $ z_{U}\approx D/2, z_{L}\approx -D/2$, so       it is only the last factor in each wave function expression that contributes to the phase shift. For the upper traverse, from $\psi_{+}$, the contribution of both plates is then $2\times(-\frac{1}{4}eD\sigma T)=-\frac{1}{2}eD\sigma T$.  Adding the contribution of the lower traverse gives  $4\times(-\frac{1}{4}eD\sigma T)=-eD\sigma T$.

It might seem from this solution that we have settled the question as to what causes the phase shift: here the phase shift is all due to the capacitor.  

However, one can also write this Hamiltonian as
\begin{equation}\label{80}
H_{\pm}=\Bigg[\frac{P_{e}^{2}}{2m}\pm\frac{1}{2}e\sigma D\Bigg]+\Bigg[\frac{P_{U}^{2}}{2M}  \pm\frac{1}{2}e\sigma \Big(Z_{U}-\frac{D}{2}\Big)\Bigg]+  \Bigg[\frac{P_{L}^{2}}{2M}  \mp\frac{1}{2}e\sigma \Big(Z_{L}+\frac{D}{2}\Big)\Bigg].
\end{equation}
In this case, the phase shift belongs completely to the electron part of the wave function, and the capacitor plate part has no phase shift associated to it (the wave function for each plate is the wave function in (\ref{79}) with additional phase factors that exactly cancel the phase that arose from the dynamics). 

Or, one could add and subtract a constant term in such a way that any fraction of the phase shift belongs to the electron wave function and the rest belongs to the capacitor.

It is clear that these are alternative ways of calculating the same phase shift, with  no reason apparent here to prefer one over the other.

\subsection{Electric Force in the Electric A-B Effect} 

In the example just discussed, the phase shift in both cases was calculated using potentials.  In one case there is the electron in the non-field-producing potential of the capacitor.  In the other case
there are the 
capacitor plates in the electric field, of the electron.  

The conceptual contrast, no force vs. force,  would be even greater if the phase shift calculation for the capacitor plates was expressed, not 
in terms of the potential, but in terms of the electron's electric force on the capacitor plates. This could not be done in the above example because the external force, 
required to separate the plates and re-unite them, intrudes.    

Accordingly, we  consider a modified  problem such that  an external force  does not act.  Instead, we suppose that, at $t=0$, the plates which coincide at $z=0$ are suddenly given velocities $\pm v_{0}$ in the $z$-direction.
They then move, under their mutual forces and the force of the electron, to  a maximum separation which we shall call $D_{\pm}$ ($D_{+}$ for the upper traverse, $D_{-}$ 
for the lower traverse), and fall back to $z=0$. 
Taking $e$ to be positive for definiteness, then $D_{+}<D_{-}$.  Therefore,  for the upper traverse, the plates return  to $z=0$ at time $T_{+}$, which is slightly earlier than the time of return $T_{-}$ for the lower electron traverse. 
 
The plates encounter a decelerating potential at $z\approx 0$ designed to slow them down to a near stop and keep them close to each other (an idea introduced by Vaidman\cite{Vaidman} in his example of the electric A-B effect) until after the electron is finally detected. Since 
the potentials are $\sim z$ and  $z\approx 0$, there is negligible phase shift contribution thereafter.  The width of the wave packets is chosen to be large enough that the positive plate wave packets from the two traverses finally overlap almost completely (similarly, of course for the negative plate wave packets), 
which is necessary if there is to be maximal interference. 

The calculation is governed by the same Hamiltonian we have already discussed, and the details of that calculation are in Appendix D.  However, they are not essential to the point be made here.  We illustrate it by considering 
the phase contributed by the positively charged plate's motion in the electric field of the electron undergoing the upper traverse using Eq.(\ref{39}),  namely,
\begin{eqnarray}\label{81}
\Phi&=&-\int_{0}^{T}dtV(z_{U}(t))=-\int_{0}^{T}dt\int_{0}^{t}dt'\frac{d}{dt'}V(z_{U}(t'))\nonumber\\
&=&-\int_{0}^{T}dt\int_{0}^{t}dt'\frac{dz_{U}(t')}{dt'}\frac{d}{dz_{U}(t')}V(z_{U}(t'))\nonumber\\
&=&\int_{0}^{T}dt\int_{0}^{t}dz_{U}(t')F(z_{U}(t')).
\end{eqnarray}
\noindent  (we have utilized $V(0)=0$) where $F(z_{U}(t)$ is the force on the capacitor plate exerted by the electron.  Thus, this phase contribution 
has been expressed in terms of the work done by the electron on the upper plate.  

\section{Concluding Remarks}

Here is a summary of the new results in this paper. 

1)  Exact solution for the magnitude and phase of a localized wave packet describing a particle moving in a vector and scalar potential corresponding to non-spatially varying forces.

2)  Approximate solution for the magnitude and phase of a particle moving in an \textit{arbitrary} vector and scalar potential.

3)  Approximate solution of the problem where both electron and solenoid are quantized, employing a variational technique, obtaining Schr\"odinger equations for electron and solenoid, showing how an \textit{extra phase} arises.

4) Application of 2) and 3) to an interference experiment, where one localized particle interacts with a second, or interacts with $N>1$ other localized particles, obtaining the magnitude and phase shift of the interference term. 
 
5) A fully quantum mechanical verification of Vaidman's semi-classical argument that the electron acting on the quantized solenoid particles gives the magnetic A-B phase shift.

6) Answer to the vexing question of why this does not result in twice the usual magnetic A-B phase shift, the phase shift from the solenoid acting on the electron added to the equal phase shift from the electron acting on the solenoid. The answer lies in 2), 3), 4): for the problem of the jointly quantized electron and solenoid, the extra phase provides the negative of these phase shifts. 

7) A time-average approach to understanding the contribution of a solenoid piece to the phase shift that gives simple intuition into the nature of the build up of the phase in the magnetic AB effect.

8) Treatment of the electric A-B effect, the exactly soluble problem of interference when a quantized electron interacts with quantized capacitor plates.  Verification of Vaidman's semi-classical argument that the electric A-B phase shift arises from the electron exerting forces on quantized capacitor plates. Showing how an extra phase is responsible for providing the electric A-B phase shift, and not twice that value.  

We have seen, in both effects, two conceptual pictures, that the A-B phase shift arises
 from the electron moving in non-field producing potentials and that
the A-B phase shift arises from the electric force exerted by the
 electron (on the solenoid particles or on the charged capacitor
plates) as was argued by Vaidman. As far as our examples are concerned,
there is no reason to prefer one of these conceptual pictures over the
other.

\appendix

\section{Wave function in Section II}\label{A}

In this appendix we do the algebra involved in obtaining the wave function for a charged object, moving on a one dimensional path, subject to a time-dependent electric field. 
\subsection{Wave function in the momentum representation}
Putting the momentum space wave function (\ref{6}),
\begin{equation}\label{A1}
\psi(p,t)\sim e^{-(\sigma^{2}+it/2m) p^{2}+\beta(t) p+i\gamma(t)-\beta_{R}^{2}(t)/4\sigma^{2}}, 
\end{equation}
\noindent with initial conditions $\gamma(0)=0$ and $\beta(0)=2p_{0}\sigma^{2}-ix_{0}$ into the momentum space Schr\"odinger equation (\ref{4}), and equating coefficients of $p$ and 1 (the coefficient of $p^{2}$ agrees on both sides of the equation) yields:
\begin{eqnarray}\label{A2}
i\frac{d}{dt}\beta&=&-\frac{q}{mc} A-2iqV' \sigma^{2}+\frac{qt}{m} V'           \nonumber\\ 
-\frac{d}{dt}\gamma&=&\frac{i}{2\sigma^{2}}\beta_{R}\frac{d}{dt}\beta_{R}+ \frac{q^{2}}{2mc^{2}}A^{2}+iqV'\beta +qg(t)\nonumber\\ 
&=&\frac{i\beta_{R}}{2\sigma^{2}}\Big[\frac{d}{dt}\beta_{R}+qV'2\sigma^{2}\Big]+\frac{q^{2}}{2mc^{2}}A^{2}-qV'\beta_{I} +qg(t)\nonumber\\
&=&\frac{q^{2}}{2mc^{2}}A^{2}-qV'\beta_{I} +qg(t)
\end{eqnarray}
\noindent where the bracketed term in the next to last equation vanishes because of the imaginary part of the first equation.
\noindent Solving for $\beta$, and putting in the initial conditions, results in
\begin{eqnarray}\label{A3}
\beta_{R}&=&2\sigma^{2}[p_{0}-qU]=2\sigma^{2}p_{cl}(t)\nonumber\\
\beta_{I}&=&-x_{0}+\frac{q}{m}[{\cal A}/c-W].
\end{eqnarray}

Now, note that 
\begin{eqnarray}\label{A4}
x_{cl}(t)&=&x_{0}+\frac{p_{0}}{m}t-\frac{q}{mc}{\cal A}-\frac{q}{m}[tU(t)-W(t)],  \nonumber\\ 
&=&\frac{t}{m}(p_{0}-qU)-\Big[-x_{0}+\frac{q}{m}({\cal A}/c-W)]=
\frac{t}{2m\sigma^{2}}\beta_{R}-\beta_{I}, 
\end{eqnarray}\label{A5}
\noindent from which follows
\begin{equation}\label{A5'}
\beta_{I}=-x_{cl}(t)+\frac{p_{cl}(t)t}{m}.
\end{equation}

In Eq.(\ref{A2}) for $\gamma$, dropping  terms quadratic in the potentials leaves only 
\begin{equation}\label{A6}
-\frac{d}{dt}\gamma=qx_{0}V'(t)+qg(t)=qV(t)-v_{0}tV'(t)\hbox{ or }\gamma=-q\int_{0}^{t}dt'V(t')+v_{0}W(t).
\end{equation}
\noindent where we have used $g(t)\equiv V(t)-(x_{0}+v_{0}t)V'(t)$.
\subsection{Wave function in the position representation} 

 The Fourier transform of (\ref{6}) is
\begin{eqnarray}\label{A7}
\psi(x,t) &\sim&\int dpe^{ipx}e^{-(\sigma^{2}+it/2m) p^{2}+\beta(t) p+i\gamma(t)-\beta_{R}^{2}(t)/4\sigma^{2}}\nonumber\\
&=&e^{\frac{(ix+\beta)^{2}}{4[\sigma^{2}+it/2m]}}e^{i\gamma}e^{-\beta_{R}^{2}(t)/4\sigma^{2}} \nonumber\\               
&=&e^{\frac{[-(x+\beta_{I})^{2}+\beta_{R}^{2}+2i\beta_{R}(x+\beta_{I})][\sigma^{2}-it/2m]-(\beta_{R}^{2}/4\sigma^{2})4[\sigma^{4}+(t/2m)^{2}]}{4[\sigma^{4}+(t/2m)^{2}]}}e^{i\gamma} .
\end{eqnarray}
\noindent We shall assume that the mass of the object is large enough so 
that there is negligible spreading of the wave packet over time $T$, so $\sigma^{4}+(t/2m)^{2}\approx \sigma^{4}$.

\subsubsection{Real part of exponent}

We have for the real part of the exponent:
\begin{eqnarray}\label{A8}
&&4\sigma^{4}\hbox{Re}\Bigg[[-(x+\beta_{I})^{2}+\beta_{R}^{2}+2i\beta_{R}(x+\beta_{I})][\sigma^{2}-it/2m]-(\beta_{R}^{2}/4\sigma^{2})4[\sigma^{4}+(t/2m)^{2}]\Bigg]\nonumber\\
&=&  -(x+\beta_{I})^{2}\sigma^{2} + 2\beta_{R}(x+\beta_{I})t/2m -(\beta_{R}t/2m\sigma)^{2}                              \nonumber\\
&=&\sigma^{2}\Big[ -x^{2}+2x[\beta_{R}(t /2m\sigma^{2})-\beta_{I}] -  [\beta_{R}(t /2m\sigma^{2})-\beta_{I}]^{2} \Big]\nonumber\\
&=&- \sigma^{2}[x-x_{cl}(t)]^{2}.
\end{eqnarray}
\noindent Thus, the mean position follows the classical trajectory.

\subsubsection{Imaginary part of exponent}
                                           
The imaginary part of the exponent, using expressions  (\ref{A3}) and (\ref{A5}) for $\beta_{R}$ and $\beta_{I}$ (and omitting $\gamma$ for the moment),  is:
\begin{eqnarray}\label{A9}   
\hbox{Imag. Part}&=&[-(x+\beta_{I})^{2}+\beta_{R}^{2}](-t/2m\sigma^{2})+2\beta_{R}(x+\beta_{I})]/4\sigma^{2}\nonumber\\
&=&\frac{t}{8m\sigma^{2}}\frac{(x-x_{cl}(t)+p_{cl}(t)t/m)^{2}}{\sigma^{2}}-\frac{p_{cl}^{2}(t)t}{2m}
+p_{cl}(t)[x-x_{cl}(t)+p_{cl}(t)t/m]\nonumber\\
&=&\frac{t}{8m\sigma^{2}}\frac{(x-x_{cl}(t)+p_{cl}(t)t/m)^{2}}{\sigma^{2}}+\frac{p_{cl}^{2}(t)t}{2m}
+p_{cl}(t)(x-x_{cl}(t))
\end{eqnarray}
\noindent That  the first term in (\ref{A9}) is negligibly smaller than the second two terms follows below, from the  negligible spread condition $ t/m\sigma^{2}<<1$ and that the real part of the exponent makes $(x-x_{cl}(t))^{2}/\sigma^{2}$ be of order 1: 
\begin{eqnarray} \label{A10}  
\hbox{Imag. Part}&=&
\frac{t}{8m\sigma^{2}}\frac{(x-x_{cl}(t))^{2}}{\sigma^{2}}+\frac{p_{cl}^{2}(t)t}{2m}\Bigg[1+\Big[\frac{t}{2m\sigma^{2}}\Big]^{2}\Bigg]+p_{cl}(t)(x-x_{cl}(t))
\Bigg[1+\Big[\frac{t}{2m\sigma^{2}}\Big]^{2}\Bigg]\nonumber\\
&&\approx\frac{t}{8m\sigma^{2}}+\frac{p_{cl}^{2}(t)t}{2m}+p_{cl}(t)(x-x_{cl}(t))\nonumber\\
&&\approx\frac{p_{cl}^{2}(t)t}{2m}+p_{cl}(t)(x-x_{cl}(t)).
\end{eqnarray}
The last approximation arises since,  
in order that the momentum of the wave packet be well defined, many oscillations of the wave function should lie within the packet width: 
\begin{equation}\label{A11} 
\frac{1}{8m\sigma^{2}}<<\frac{p_{cl}^{2}(t)}{2m}=\frac{k_{cl}^{2}}{2m}.
\end{equation}

Combining the exponents in (\ref{A6}), (\ref{A8}) and (\ref{A10}), Wwe have finally arrived at the wave packet expression:
\begin{equation}\label{A12}
\psi(x,t)=Ne^{-\frac{(x-x_{cl}(t))^{2}}{4\sigma^{2}}}e^{ip_{cl}(t)(x-x_{cl}(t))}e^{ip_{cl}^{2}(t)t/2m}e^{-iq\int_{0}^{t}dt'V(t')+iv_{0}W(t)}.
\end{equation}

\section{General Case Phase of Section III Reduces to Special Case Phase of Section II For Solely Time-Dependent Electric Field}\label{B}

If the force depends only upon time, i.e.,  if
$V'(x,t)=V'(t)$, $A(x,t)=A(t)$, the phase angle is given by  Eq. (\ref{8}) in Section II: 
\begin{equation}\label{B1}
\theta(x,t)=p_{cl}(t)[x-x_{cl}(t)]+\frac{t}{2m}p_{cl}^{2}(t)-q\int_{0}^{t}dt'V(t')+v_{0}qW(t),
\end{equation}
 \noindent (neglecting some terms squared in the potentials (i.e., $\sim q^{2}$), which is the order we employ).  
We wish to show here that the phase expression (\ref{30}) in Section III, 
\begin{equation}\label{B2}
\theta(x,t)=p_{cl}(t)[x-x_{cl}(t)]+\int_{0}^{t}dt'\frac{1}{2m}p_{cl}^{2}(t')-q\int_{0}^{t}dt'V(x_{cl}(t'),t').
\end{equation}
\noindent for the general case of the electric field depending  on position and time reduces to (\ref{B1}) when there is only time-dependence.

  In that case,
\begin{eqnarray}\label{B3}
 p_{cl}(t)&=&mv_{cl}(t)+\frac{q}{c}A(x_{cl}(t),t)\rightarrow mv_{cl}(t)+\frac{q}{c}A(t)=mv_{0}-U(t)\nonumber\\
        V(x_{cl}(t'),t')&\rightarrow &V(t').
\end{eqnarray}
\noindent (The latter is the zeroth order term in the expansion of the scalar potential $V(x,t)$ in $x-x_{cl}(t)$)  Then we have:
\begin{eqnarray}\label{B4}
\int_{0}^{t}dt'\frac{1}{2m}p_{cl}^{2}(t')&=&\int_{0}^{t}dt'\frac{1}{2m}[mv_{0}-qU(t')]^{2}=\frac{m}{2}v_{0}^{2}t-v_{0}q\int_{0}^{t}dt'U(t')\nonumber\\
&=&\frac{m}{2}v_{0}^{2}t-v_{0}q[tU(t)-W(t)]=\frac{1}{2m}p_{cl}^{2}(t)t+v_{0}qW(t).
\end{eqnarray}
\noindent When (\ref{B4}) is inserted into (\ref{B2}), we obtain (\ref{B1}).

\section{Calculations with explication involved in time average approach}\label{C}

In section VD, we argued that one gets the correct contribution to the phase associated to the motion of a piece, subjected to the vector potential of the electron, by 
recognizing that,  effectively,  there is always a piece  at angle $\phi$ that feels the time-averaged vector potential:
\begin{eqnarray}\label{C1}
\langle A\rangle_{T}&\equiv &\frac{1}{T}\int_{0}^{T}dtA(t)= \frac{1}{T}\int_{0}^{T}dt\frac{e}{cS}\hat{\boldsymbol\phi}\cdot{\bf u}\nonumber\\
&=&\frac{u}{\pi R}\int_{0}^{T}dt(\pm u) e\Big[\frac{\cos(\phi-\phi'(t))}{cD}+\frac{aR\cos^{2}(\phi-\phi'(t))}{cD^{3}}\Big]\nonumber\\
&=&\pm\frac{eu}{cD}\Big[\frac{2}{\pi}\cos\phi +\frac{aR}{2D^{2}}\Big]
\end{eqnarray}
\noindent where $S\equiv\sqrt{a^{2}+R^{2}+z^{2}-2aR\cos(\phi-\phi')}$  has been expanded to first order in $a$, $D\equiv \sqrt{R^{2}+z^{2}}$, $uT=\pi R$ has been employed, as has $d\phi'(t)=\pm\frac{u}{R}dt $ (+ for right traverse where 
$\phi'(t)=\frac{u}{R}t-\frac{\pi}{2}$, - for left traverse where $\phi'(t)=-\frac{u}{R}t-\frac{\pi}{2}$).

Now, since the vector potential is constant along the path of the piece, and noting that the distance traveled by each piece during $T$ is 
$v_{0}T=v_{0}\frac{\pi R}{u}$,   the phase contributed by a piece at angle $\phi$ is given in Eq.(\ref{73}), 
\begin{eqnarray}\label{C2}
\Lambda_{\pm}&=&\int_{path}ds\frac{q}{c}\langle A\rangle_{T}\nonumber\\
&=&\pm v_{0}\frac{eq\pi R}{c^{2}D}\Big[\frac{2}{\pi}\cos\phi +\frac{aR}{2D^{2}}\Big],
\end{eqnarray}
Integrating over all the pieces using $q$ in Eq.(\ref{64}) gives for the total phase for a path and a cylinder:
\begin{eqnarray}\label{C3}
\hbox{Phase}_{\pm}&=&Q\int_{0}^{2\pi}\frac{d\phi}{2\pi}\int_{-L/2}^{L/2}\frac{dz}{L}\Lambda_{\pm}\nonumber\\
&=&Q\int_{0}^{2\pi}\frac{d\phi}{2\pi}\int_{-L/2}^{L/2}\frac{dz}{L}\frac{1}{D^{3}}(\pm) v_{0}\frac{eq\pi a R^{2}}{2c^{2}}=\pm\frac{Qev_{0}a\pi}{Lc^{2}}.
\end{eqnarray}

The phase difference between the two paths gives twice the magnitude of this.  Accounting for both cylinders requires another factor of 2, resulting in the standard A-B phase shift. 

This result is the same as obtained in section VC, where the integral over angle was considered first, followed by the integrals over time: here the integrals are taken in reverse order.

\section{Analysis of the phase shift in the electric A-B effect for the setup discussed in Section  VIC}\label{D}

To calculate the phase shift with the Hamiltonian (\ref{78}), we must find the classical motion of the electron and the capacitor plates.  
The electron feels no force, and the plates feel a constant force with the potential energy $\frac{\sigma}{2}[\sigma A\pm e][z_{U}-z_{L}]$, 
where $z_{U}, z_{L}$ are the upper and lower capacitor plate positions, The upper sign in $\pm$ refers to the electron's traverse above the capacitor (trajectory $A$), 
the lower sign refers to the electron's traverse below the capacitor (trajectory $B$).

The solutions of the equations of motion are, for the electron and plates:
\begin{eqnarray}\label{D1}
v_{e}&=&u, \quad x_{e}=ut\nonumber\\
v_{U}&=&-v_{L}=v_{0}-\frac{\sigma}{2M}[\sigma A\pm e]t, \quad z_{U}=-z_{L}= v_{0}t-\frac{\sigma}{4M}[\sigma A\pm e]t^{2}. 
\end{eqnarray}
\noindent 

The largest separation of the plates for the two possible traverses, and the time it takes  the plates to return to $z=0$, are
\begin{equation}\label{D2}
D_{\pm}= \frac{1}{4}v_{0}T_{\pm}, \quad T_{\pm}=\frac{4Mv_{0}}{\sigma[\sigma A\pm e]}.
\end{equation}
\noindent Taking  $e$ to be positive for simplicity, then $T_{+}<T_{-}$.  We shall take $e<<\sigma A$, so that time difference is small, but we shall see that nonetheless it has an important effect.  

The phase shift is then 
\begin{eqnarray}\label{D3}
\Phi^{A}-\Phi^{B}&\equiv&-\int_{0}^{T }dt[V^{A}(t)-V^{B}(t)]=-\int_{0}^{T_{+}}dt\frac{\sigma}{2}[\sigma A+ e][z_{U}(t)-z_{L}(t)]_{+}+\-\int_{0}^{T_{-}}dt\frac{\sigma}{2}[\sigma A+ e][z_{U}(t)-z_{L}(t)]_{-}\nonumber\\
&=&-\frac{\sigma e v_{0}}{6}[T_{-}^{2}+T_{+}^{2}]+\frac{\sigma^{2}Av_{0}}{6}[T_{-}^{2}-T_{+}^{2}]\nonumber\\
&\approx& -\frac{1}{3}\sigma e v_{0}{\bar T}^{2}+\frac{2}{3}\sigma e v_{0}{\bar T}^{2}=\frac{1}{3}\sigma e v_{0}{\bar T}^{2}.
\end{eqnarray}
\noindent where we have written ${\bar T}\equiv\frac{4Mv_{0}}{\sigma^{2}A}$ and the approximation is to drop terms $\sim e/\sigma A$ compared to 1. 

The first term in the second line of (\ref{D3}) is the negative phase shift due to the plates moving in the potential of the electron.  The second term in that line 
is a positive phase shift due to one plate moving in the potential of the other. That phase shift would vanish if the plates spent the same amount of time on their 
trajectories when the electron was above them as below them.  However, that is not the case, due to the force of the electron on the plates.  

Perhaps surprisingly, the 
net phase shift is positive.  This is because, although the time difference in travel is relatively small, the self-potential energy of the plates is relatively large compared to their potential energy 
in the field of the electron, and the phase shift due to the larger potential energy wins out. 

Again, as in section VIB, although it appears here that the phase shift is totally due to the motion of the plates under the force of the electron, one can add and subtract a term from the 
Hamiltonian (\ref{78}), rewriting it as
\begin{equation}\label{D4}
H_{\pm}=\Bigg[\frac{P^{2}}{2m}+\frac{\sigma}{2}[\sigma A\pm e][z_{U}(t)-z_{L}(t)]_{\pm}\Bigg]+\Bigg[\frac{P_{U}^{2}}{2M} + \frac{\sigma}{2}[\sigma A\pm e][Z_{U}-z_{U}(t)]_{\pm}\Bigg]+  
\Bigg[\frac{P_{L}^{2}}{2M} - \frac{\sigma}{2}[\sigma A\pm e][Z_{L}-z_{L}(t)]_{\pm}\Bigg].
\end{equation}
Now the electron part of the Hamiltonian has in it the time varying, spatially constant potential it sees, and this is completely responsible for the phase shift, while the capacitor plate part of the Hamiltonian 
makes no contribution to the phase shift.

\acknowledgments{ We would like to thank Lev Vaidman for very helpful conversations. We are also very grateful to 
an anonymous referee who insisted our previous assertion, that the interacting quantized electron and solenoid results in  
twice the A-B phase shift, was wrong, leading us to greater effort, resulting in the resolution in this paper.}

 \end{document}